\newif\ifOL
\OLfalse

\documentclass[5p,authoryear]{elsarticle}
\journal{Imaging Neuroscience}

\newcommand{\Papertitle}{Unconstrained quantitative magnetization transfer imaging: disentangling $T_1$ of the free and semi-solid spin pools}

\newif\ifOL
\OLfalse

\ifOL
\else
\usepackage{animate}
\usepackage{tikz}
\usepackage{pgfplots}

\usetikzlibrary{spy,backgrounds,arrows,decorations.pathmorphing,backgrounds,positioning,fit,matrix,calc,shapes.multipart,shapes.misc,pgfplots.statistics, pgfplots.colorbrewer}

\pgfplotsset{compat=1.18,
	boxplot/estimator=R1,
	colormap={blackwhite}{gray(0cm)=(0); gray(1cm)=(1)},
	colormap={parula}{
			rgb255=(53,42,135)
			rgb255=(15,92,221)
			rgb255=(18,125,216)
			rgb255=(7,156,207)
			rgb255=(21,177,180)
			rgb255=(89,189,140)
			rgb255=(165,190,107)
			rgb255=(225,185,82)
			rgb255=(252,206,46)
			rgb255=(249,251,14)},
	colormap={mybluered}{
			rgb255(0cm)=(0,0,180)
			rgb255(1cm)=(0,180,180)
			rgb255(2cm)=(70,180,0)
			rgb255(3cm)=(180,180,0)
			rgb255(4cm)=(255,0,0)
			rgb255(5cm)=(128,0,0)},
	colormap={mybluewhitered_m10_2}{
			rgb255(0cm)=(0,0,255)
			rgb255(10cm)=(255,255,255)
			rgb255(12cm)=(255,0,0)},
	colormap={mybluewhitered_m1_2}{
			rgb255(0cm)=(0,0,255)
			rgb255(1cm)=(255,255,255)
			rgb255(3cm)=(255,0,0)},
	colormap={mybluewhitered_0_1}{
			rgb255(0cm)=(255,255,255)
			rgb255(1cm)=(255,0,0)},
	colormap={gist_earth}{
			rgb255=(0.0,0.0,0.0)
			rgb255=(1.7085000000000001,0.0,62.424)
			rgb255=(3.3914999999999997,0.0,89.8365)
			rgb255=(5.1000000000000005,1.4789999999999999,113.985)
			rgb255=(6.8085,7.343999999999999,116.4075)
			rgb255=(8.4915,13.209,116.8665)
			rgb255=(10.200000000000001,19.074,117.3255)
			rgb255=(11.9085,24.939,117.7845)
			rgb255=(13.5915,30.804000000000002,118.2435)
			rgb255=(15.299999999999999,36.669000000000004,118.7025)
			rgb255=(16.983,42.534,119.1615)
			rgb255=(18.6915,48.3735,119.6205)
			rgb255=(20.400000000000002,53.677499999999995,120.10499999999999)
			rgb255=(22.083,58.981500000000004,120.564)
			rgb255=(23.7915,64.2855,121.02300000000001)
			rgb255=(25.5,69.58949999999999,121.482)
			rgb255=(27.183,74.8935,121.941)
			rgb255=(28.8915,79.917,122.39999999999999)
			rgb255=(30.599999999999998,84.6855,122.859)
			rgb255=(32.282999999999994,89.454,123.3435)
			rgb255=(33.9915,94.2225,123.8025)
			rgb255=(35.7,98.838,124.2615)
			rgb255=(37.383,102.86699999999999,124.7205)
			rgb255=(39.091499999999996,106.896,125.1795)
			rgb255=(40.774499999999996,110.925,125.63850000000001)
			rgb255=(42.483,114.954,126.0975)
			rgb255=(44.191500000000005,118.983,126.5565)
			rgb255=(45.8745,123.012,127.041)
			rgb255=(47.583,127.041,127.5)
			rgb255=(48.909,129.13199999999998,125.23049999999999)
			rgb255=(50.1075,130.5855,122.1195)
			rgb255=(51.306,132.0645,118.983)
			rgb255=(52.5045,133.518,115.872)
			rgb255=(53.7285,134.99699999999999,112.7355)
			rgb255=(54.927,136.476,109.6245)
			rgb255=(56.125499999999995,137.92950000000002,106.51350000000001)
			rgb255=(57.324,139.4085,103.377)
			rgb255=(58.5225,140.862,100.266)
			rgb255=(59.721,142.341,97.12950000000001)
			rgb255=(60.945,143.7945,94.0185)
			rgb255=(62.1435,145.27349999999998,90.882)
			rgb255=(63.342000000000006,146.7525,87.771)
			rgb255=(64.5405,148.20600000000002,84.66000000000001)
			rgb255=(65.73899999999999,149.685,81.5235)
			rgb255=(66.9375,151.1385,78.4125)
			rgb255=(68.136,152.6175,75.27600000000001)
			rgb255=(69.8955,154.071,72.16499999999999)
			rgb255=(75.5565,155.54999999999998,70.5585)
			rgb255=(81.2175,157.029,72.063)
			rgb255=(86.904,158.48250000000002,73.542)
			rgb255=(92.565,159.9615,75.021)
			rgb255=(98.226,161.415,76.5)
			rgb255=(103.887,162.894,78.00450000000001)
			rgb255=(109.57350000000001,164.06699999999998,79.48349999999999)
			rgb255=(115.23450000000001,165.18900000000002,80.9625)
			rgb255=(120.74249999999999,166.311,82.11)
			rgb255=(124.95,167.4075,82.926)
			rgb255=(129.1575,168.5295,83.742)
			rgb255=(133.365,169.6515,84.5325)
			rgb255=(137.5725,170.77349999999998,85.3485)
			rgb255=(141.78,171.8955,86.16449999999999)
			rgb255=(145.9875,173.01749999999998,86.95500000000001)
			rgb255=(150.195,174.1395,87.771)
			rgb255=(154.4025,175.2615,88.5615)
			rgb255=(158.60999999999999,176.3835,89.3775)
			rgb255=(162.792,177.48,90.1935)
			rgb255=(166.9995,178.602,90.984)
			rgb255=(171.207,179.724,91.8)
			rgb255=(175.41449999999998,180.846,92.59049999999999)
			rgb255=(179.622,181.968,93.40650000000001)
			rgb255=(183.192,182.42700000000002,94.2225)
			rgb255=(184.263,180.2085,95.01299999999999)
			rgb255=(185.334,178.0155,95.82900000000001)
			rgb255=(186.405,175.8225,96.6195)
			rgb255=(187.476,173.60399999999998,97.4355)
			rgb255=(188.5215,171.411,98.2515)
			rgb255=(189.5925,169.218,99.042)
			rgb255=(190.6635,166.9995,99.858)
			rgb255=(191.7345,164.8065,100.6485)
			rgb255=(193.1115,163.2255,104.5755)
			rgb255=(195.9675,164.3985,110.823)
			rgb255=(198.8235,165.75,117.0705)
			rgb255=(201.67950000000002,167.94299999999998,123.318)
			rgb255=(204.5355,170.1615,129.5655)
			rgb255=(207.366,172.38000000000002,135.813)
			rgb255=(210.222,174.573,142.06050000000002)
			rgb255=(213.078,176.7915,148.308)
			rgb255=(215.934,179.469,154.5555)
			rgb255=(218.79,183.141,160.80300000000003)
			rgb255=(221.646,186.66,167.0505)
			rgb255=(224.4765,190.4595,174.1395)
			rgb255=(227.33249999999998,195.45749999999998,181.815)
			rgb255=(230.18849999999998,200.43,189.516)
			rgb255=(233.0445,205.428,197.1915)
			rgb255=(235.90050000000002,210.42600000000002,204.89249999999998)
			rgb255=(238.75650000000002,216.1635,212.568)
			rgb255=(241.6125,222.0285,220.2435)
			rgb255=(244.443,228.7605,227.9445)
			rgb255=(247.299,236.2065,235.62)
			rgb255=(250.155,243.6525,243.32100000000003)
			rgb255=(253.011,250.9965,250.9965)
		},
	colormap={berlin}{
			rgb255=(158.37591,175.99641,254.87428500000001)
			rgb255=(156.100035,175.75313999999997,253.82037)
			rgb255=(153.81651,175.50375,252.765945)
			rgb255=(151.521,175.250535,251.70846)
			rgb255=(149.21707500000002,174.99324,250.64511)
			rgb255=(146.90244,174.73161,249.57691499999999)
			rgb255=(144.57505500000002,174.46233,248.50361999999998)
			rgb255=(142.236705,174.18999,247.42344)
			rgb255=(139.89045000000002,173.90796,246.33408)
			rgb255=(137.52838500000001,173.619045,245.23452)
			rgb255=(135.15867,173.32120500000002,244.12425000000002)
			rgb255=(132.775185,173.01342,243.00020999999998)
			rgb255=(130.38022500000002,172.69365,241.85883)
			rgb255=(127.97506499999999,172.35654,240.70036499999998)
			rgb255=(125.55384,172.00413,239.520735)
			rgb255=(123.12216,171.63412499999998,238.31637)
			rgb255=(120.675945,171.24015,237.086505)
			rgb255=(118.22055,170.82399,235.82553000000001)
			rgb255=(115.752405,170.37876,234.532425)
			rgb255=(113.274315,169.90012499999997,233.20209000000003)
			rgb255=(110.7822,169.38910499999997,231.82968000000002)
			rgb255=(108.284475,168.84060000000002,230.41443)
			rgb255=(105.77859000000001,168.246705,228.950475)
			rgb255=(103.268625,167.608695,227.43577499999998)
			rgb255=(100.759935,166.917645,225.86625)
			rgb255=(98.25048000000001,166.17686999999998,224.23884)
			rgb255=(95.750715,165.37668,222.553035)
			rgb255=(93.26421,164.51682,220.805265)
			rgb255=(90.795045,163.59576,218.994255)
			rgb255=(88.345515,162.61146000000002,217.120005)
			rgb255=(85.93041,161.56136999999998,215.18302500000001)
			rgb255=(83.54871,160.443195,213.184335)
			rgb255=(81.214185,159.260505,211.12393500000002)
			rgb255=(78.932445,158.012535,209.00514)
			rgb255=(76.69992,156.696735,206.83254000000002)
			rgb255=(74.538795,155.324325,204.606645)
			rgb255=(72.44499,153.888675,202.33485)
			rgb255=(70.43227499999999,152.39667,200.01843)
			rgb255=(68.49172499999999,150.8529,197.661465)
			rgb255=(66.63354,149.26042500000003,195.2739)
			rgb255=(64.86384,147.62153999999998,192.85548)
			rgb255=(63.177015,145.94364,190.413345)
			rgb255=(61.57332,144.22698,187.95183)
			rgb255=(60.056325,142.48074,185.474505)
			rgb255=(58.60971,140.70951,182.988)
			rgb255=(57.248265,138.91125,180.490275)
			rgb255=(55.968675000000005,137.09514,177.98949)
			rgb255=(54.74697,135.26041500000002,175.48845)
			rgb255=(53.59386,133.414215,172.986135)
			rgb255=(52.501695,131.55373500000002,170.48739)
			rgb255=(51.451605,129.69249,167.990685)
			rgb255=(50.45889,127.82078999999999,165.50265)
			rgb255=(49.51386,125.94526499999999,163.01818500000002)
			rgb255=(48.59178,124.067955,160.54086)
			rgb255=(47.71356,122.19115500000001,158.07398999999998)
			rgb255=(46.85676,120.31563000000001,155.611455)
			rgb255=(46.027499999999996,118.44087,153.15861)
			rgb255=(45.228075,116.564835,150.714435)
			rgb255=(44.43732,114.69594000000001,148.278165)
			rgb255=(43.662119999999994,112.83086999999999,145.85133)
			rgb255=(42.90171,110.96886,143.43393)
			rgb255=(42.14946,109.11041999999999,141.020355)
			rgb255=(41.421945,107.25504,138.618765)
			rgb255=(40.683975,105.40935,136.22355)
			rgb255=(39.968444999999996,103.56748499999999,133.83828)
			rgb255=(39.245774999999995,101.72766,131.45862)
			rgb255=(38.53611,99.89803500000001,129.08865)
			rgb255=(37.828230000000005,98.070705,126.72786)
			rgb255=(37.138455,96.25281000000001,124.376505)
			rgb255=(36.434145,94.43746499999999,122.02872)
			rgb255=(35.73519,92.630535,119.694195)
			rgb255=(35.05383,90.83202,117.360945)
			rgb255=(34.368135,89.03630999999999,115.042485)
			rgb255=(33.680145,87.24825,112.731675)
			rgb255=(32.997254999999996,85.469115,110.42571)
			rgb255=(32.317425,83.68972500000001,108.12918)
			rgb255=(31.642950000000003,81.921045,105.84310500000001)
			rgb255=(30.97128,80.158485,103.56672)
			rgb255=(30.319245000000002,78.4023,101.29467)
			rgb255=(29.66058,76.65504,99.03588)
			rgb255=(29.001405,74.91415500000001,96.78423)
			rgb255=(28.352684999999997,73.1799,94.54074)
			rgb255=(27.696315000000002,71.455335,92.31102)
			rgb255=(27.070545,69.74173499999999,90.0864)
			rgb255=(26.43585,68.02788000000001,87.87147)
			rgb255=(25.801665,66.32754,85.66775999999999)
			rgb255=(25.18788,64.634085,83.47221)
			rgb255=(24.568485,62.94675,81.289665)
			rgb255=(23.985045,61.26732,79.11808500000001)
			rgb255=(23.405939999999998,59.600384999999996,76.94829)
			rgb255=(22.82403,57.947475,74.79609)
			rgb255=(22.271955000000002,56.304,72.65307)
			rgb255=(21.711209999999998,54.6618,70.52687999999999)
			rgb255=(21.182595000000003,53.035155,68.403495)
			rgb255=(20.67999,51.415905,66.29796)
			rgb255=(20.178150000000002,49.817055,64.204155)
			rgb255=(19.707929999999998,48.22968,62.115195)
			rgb255=(19.270605,46.650465,60.05301)
			rgb255=(18.868215,45.092925,57.99567)
			rgb255=(18.46455,43.550174999999996,55.955414999999995)
			rgb255=(18.116474999999998,42.022725,53.9325)
			rgb255=(17.790585,40.519754999999996,51.92514)
			rgb255=(17.49759,39.025200000000005,49.933589999999995)
			rgb255=(17.2278,37.566345,47.97162)
			rgb255=(16.999575,36.126104999999995,46.01526)
			rgb255=(16.810364999999997,34.69938,44.093835)
			rgb255=(16.661444999999997,33.327225,42.19179)
			rgb255=(16.552305,31.963994999999997,40.333095)
			rgb255=(16.48218,30.63366,38.49123)
			rgb255=(16.451835000000003,29.342850000000002,36.691694999999996)
			rgb255=(16.46127,28.10661,34.924034999999996)
			rgb255=(16.510995,26.883885000000003,33.208650000000006)
			rgb255=(16.6005,25.716495,31.510095)
			rgb255=(16.672665,24.599595,29.878349999999998)
			rgb255=(16.721369999999997,23.54619,28.30704)
			rgb255=(16.80246,22.491255,26.770410000000002)
			rgb255=(16.92894,21.45417,25.31844)
			rgb255=(17.11254,20.413005,23.926395)
			rgb255=(17.389215,19.405245,22.559849999999997)
			rgb255=(17.7786,18.432165,21.171375)
			rgb255=(18.267944999999997,17.506770000000003,19.77372)
			rgb255=(18.86439,16.58979,18.38805)
			rgb255=(19.53198,15.722534999999999,16.996005)
			rgb255=(20.307434999999998,14.93025,15.588915)
			rgb255=(21.155565,14.19483,14.214975)
			rgb255=(22.066935,13.514235000000001,12.83568)
			rgb255=(23.030325,12.928245,11.485199999999999)
			rgb255=(24.0363,12.432015,10.142115)
			rgb255=(25.071345,11.995455,8.844165)
			rgb255=(26.12679,11.63412,7.66887)
			rgb255=(27.216659999999997,11.399775,6.63306)
			rgb255=(28.30143,11.21286,5.706645)
			rgb255=(29.387475,11.11698,4.88325)
			rgb255=(30.484485,11.109585000000001,4.156245)
			rgb255=(31.57206,11.184555,3.5182350000000002)
			rgb255=(32.666775,11.337045,2.9549399999999997)
			rgb255=(33.740325,11.533394999999999,2.430405)
			rgb255=(34.795004999999996,11.77182,2.013225)
			rgb255=(35.861925,12.08037,1.65801)
			rgb255=(36.945420000000006,12.40167,1.358385)
			rgb255=(38.05365,12.708179999999999,1.10823)
			rgb255=(39.189674999999994,13.004235,0.901935)
			rgb255=(40.368795,13.29315,0.73491)
			rgb255=(41.56857,13.570590000000001,0.602565)
			rgb255=(42.791805,13.831199999999999,0.5005649999999999)
			rgb255=(44.04768,14.068859999999999,0.425595)
			rgb255=(45.339254999999994,14.28459,0.374595)
			rgb255=(46.630065,14.4891,0.3417)
			rgb255=(47.95479,14.68137,0.32181000000000004)
			rgb255=(49.274415,14.921069999999999,0.31263)
			rgb255=(50.608065,15.18525,0.312885)
			rgb255=(51.96339,15.427755,0.32130000000000003)
			rgb255=(53.31846,15.67893,0.33711)
			rgb255=(54.68985,15.991050000000001,0.36006)
			rgb255=(56.073735,16.274865000000002,0.389895)
			rgb255=(57.462975,16.581885,0.42712500000000003)
			rgb255=(58.86828,16.905735,0.472515)
			rgb255=(60.28761,17.249475,0.52734)
			rgb255=(61.71408,17.61846,0.5928749999999999)
			rgb255=(63.158655,17.966790000000003,0.67116)
			rgb255=(64.61445,18.35643,0.76449)
			rgb255=(66.08988000000001,18.7782,0.8759250000000001)
			rgb255=(67.574235,19.185435,1.0085250000000001)
			rgb255=(69.08817,19.626075,1.165605)
			rgb255=(70.61664,20.099610000000002,1.351755)
			rgb255=(72.169335,20.608845,1.5710549999999999)
			rgb255=(73.74498,21.134145,1.828605)
			rgb255=(75.34383,21.694125,2.128995)
			rgb255=(76.97379000000001,22.3023,2.48013)
			rgb255=(78.63868500000001,22.92756,2.9210249999999998)
			rgb255=(80.33112000000001,23.59515,3.3976200000000003)
			rgb255=(82.06053,24.32496,3.930315)
			rgb255=(83.82818999999999,25.09098,4.5339)
			rgb255=(85.63027500000001,25.902900000000002,5.214494999999999)
			rgb255=(87.47418,26.769135,5.9772)
			rgb255=(89.35531499999999,27.7032,6.826605)
			rgb255=(91.27648500000001,28.70382,7.76628)
			rgb255=(93.235395,29.74779,8.815605)
			rgb255=(95.23485000000001,30.847604999999998,9.974324999999999)
			rgb255=(97.273065,32.02953,11.141715000000001)
			rgb255=(99.343665,33.266535,12.360105)
			rgb255=(101.450475,34.54587,13.54968)
			rgb255=(103.5861,35.902725000000004,14.75124)
			rgb255=(105.74595000000001,37.29987,15.992325000000001)
			rgb255=(107.923395,38.754645000000004,17.259674999999998)
			rgb255=(110.118435,40.26603,18.62622)
			rgb255=(112.31322,41.82714,20.048099999999998)
			rgb255=(114.516675,43.418595,21.58422)
			rgb255=(116.71452,45.04983,23.171595)
			rgb255=(118.91007,46.719314999999995,24.820425)
			rgb255=(121.09949999999999,48.42144,26.53632)
			rgb255=(123.27210000000001,50.152635,28.314944999999998)
			rgb255=(125.43705,51.89658,30.138450000000002)
			rgb255=(127.58211000000001,53.660415,32.002755)
			rgb255=(129.71595,55.443375,33.910664999999995)
			rgb255=(131.829135,57.23016,35.858865)
			rgb255=(133.92523500000001,59.043465,37.828995)
			rgb255=(136.003995,60.84912,39.846555)
			rgb255=(138.0672,62.667525,41.879414999999995)
			rgb255=(140.117655,64.495365,43.927575)
			rgb255=(142.15281000000002,66.323205,46.002765000000004)
			rgb255=(144.17139,68.150025,48.1032)
			rgb255=(146.18436,69.987555,50.215619999999994)
			rgb255=(148.18356,71.826615,52.335435000000004)
			rgb255=(150.1746,73.66797,54.474375)
			rgb255=(152.16258000000002,75.50907,56.62377)
			rgb255=(154.14342,77.352975,58.784895000000006)
			rgb255=(156.11814,79.207335,60.958259999999996)
			rgb255=(158.09387999999998,81.056085,63.14259)
			rgb255=(160.06554,82.90866,65.328195)
			rgb255=(162.03669,84.772965,67.527825)
			rgb255=(164.009115,86.63497500000001,69.73995000000001)
			rgb255=(165.983835,88.50131999999999,71.955645)
			rgb255=(167.95855500000002,90.37072500000001,74.176185)
			rgb255=(169.936845,92.246505,76.40820000000001)
			rgb255=(171.91947,94.12585499999999,78.645825)
			rgb255=(173.903625,96.01209,80.890845)
			rgb255=(175.894665,97.89959999999999,83.14096500000001)
			rgb255=(177.88698,99.79526999999999,85.40689499999999)
			rgb255=(179.88567,101.69247,87.66798)
			rgb255=(181.88844,103.599105,89.94359999999999)
			rgb255=(183.89529,105.50803499999999,92.22381)
			rgb255=(185.90877,107.42104499999999,94.50759000000001)
			rgb255=(187.92684,109.34298,96.80208)
			rgb255=(189.9495,111.26899499999999,99.102945)
			rgb255=(191.97700500000002,113.200365,101.4084)
			rgb255=(194.011905,115.13556000000001,103.72048500000001)
			rgb255=(196.049355,117.07661999999999,106.03869)
			rgb255=(198.09521999999998,119.02278,108.36531000000001)
			rgb255=(200.144145,120.973785,110.69345999999999)
			rgb255=(202.198425,122.9304,113.033595)
			rgb255=(204.257295,124.889565,115.378575)
			rgb255=(206.32305,126.85893,117.72636000000001)
			rgb255=(208.39161000000001,128.82778499999998,120.08205)
			rgb255=(210.46629000000001,130.80531000000002,122.44335)
			rgb255=(212.546835,132.78665999999998,124.808475)
			rgb255=(214.63146,134.774385,127.184565)
			rgb255=(216.720675,136.765425,129.56448)
			rgb255=(218.81346,138.761055,131.94924)
			rgb255=(220.91262,140.763315,134.340375)
			rgb255=(223.01484,142.769145,136.73559)
			rgb255=(225.121395,144.779055,139.13896499999998)
			rgb255=(227.23305,146.79585,141.54591)
			rgb255=(229.348785,148.814175,143.96025)
			rgb255=(231.46758,150.84015,146.37969)
			rgb255=(233.587905,152.87046,148.80576)
			rgb255=(235.71384,154.905615,151.239225)
			rgb255=(237.84207,156.9423,153.67549499999998)
			rgb255=(239.97438,158.988165,156.118395)
			rgb255=(242.10898500000002,161.03556,158.56716)
			rgb255=(244.24716,163.088565,161.024085)
			rgb255=(246.38584500000002,165.14514,163.48381500000002)
			rgb255=(248.527845,167.20809,165.95196)
			rgb255=(250.67264999999998,169.274865,168.42418500000002)
			rgb255=(252.81924,171.34444499999998,170.90508)
			rgb255=(254.967615,173.41836,173.38725000000002)
		}
}

\pgfdeclarelayer{background}
\pgfdeclarelayer{foreground}
\pgfsetlayers{background,main,foreground}

\usepgfplotslibrary{external,fillbetween}
\RequirePackage{shellesc}
\fi

\usepackage{xcolor}
\definecolor{UKLred} {RGB}{207, 25,  59}
\definecolor{UKLblue}{RGB}{ 47, 63, 157}
\definecolor{NYUpurple} {RGB}{88,15,139}
\definecolor{Pastrami} {RGB}{229,85,79}
\definecolor{TheLake}{RGB}{72,159,223}
\definecolor{EastRiver}{RGB}{0,115,152}
\definecolor{SpicyMustard}{RGB}{203,160,82}
\definecolor{CentralPark}{RGB}{0,108,91}
\definecolor{ProspectPark}{RGB}{64,192,172}
\definecolor{turquois}{rgb}{0,0.75,0.75}%

\usepackage{colortbl}

\usepackage{amsmath}
\usepackage{amssymb}
\usepackage{amsthm}
\usepackage{amsfonts}
\usepackage{upgreek}
\usepackage{mathabx}
\usepackage{xfrac}

\usepackage{widetext}

\usepackage{subcaption}

\usepackage{xr}

\usepackage{nameref}
\usepackage{hyperref}
\newcommand{\linkcolor}{blue}
\hypersetup{
	pdfpagelayout=TwoPageLeft,
	pdfstartview=Fit,
	bookmarksopen=false,
	pdftitle={\Papertitle},
	pdfauthor={Jakob Assl\"ander},
	pdfsubject={Manuscript},
	breaklinks=true,
	colorlinks=true,
	linkcolor=\linkcolor,
	anchorcolor=black,
	citecolor=\linkcolor,
	filecolor=black,
	menucolor=red,
	urlcolor=\linkcolor,
	pdfencoding=auto
}

\makeatletter
\newcommand*{\addFileDependency}[1]{
	\typeout{(#1)}
	\@addtofilelist{#1}
	\IfFileExists{#1}{}{\typeout{No file #1.}}
}
\makeatother
\newcommand*{\myexternaldocument}[1]{%
	\externaldocument{#1}%
	\addFileDependency{#1.tex}%
	\addFileDependency{#1.aux}%
}
\myexternaldocument{./Supporting_Information}

\begin{document}

\begin{frontmatter}
	\title{\Papertitle}

	\author[1,2]{Jakob Assl\"ander\footnotemark\textsuperscript{,}}
	\author[1,2,3]{Andrew Mao}
	\author[1,2]{Elisa Marchetto}
	\author[4]{Erin S Beck}
	\author[4]{Francesco La Rosa}
	\author[5]{Robert W Charlson}
	\author[1]{Timothy M Shepherd}
	\author[1,2]{Sebastian Flassbeck}

	\affiliation[1]{
		organization={Center for Biomedical Imaging, Dept. of Radiology, New York University School of Medicine},
		addressline={650 1st Avenue},
		city={New York},
		postcode={10016},
		state={NY},
		country={USA}
	}

	\affiliation[2]{
		organization={Center for Advanced Imaging Innovation and Research (CAI$^2$R), Dept. of Radiology, New York University School of Medicine},
		addressline={650 1st Avenue},
		city={New York},
		postcode={10016},
		state={NY},
		country={USA}}

	\affiliation[3]{
		organization={Vilcek Institute of Graduate Biomedical Sciences, New York University School of Medicine},
		addressline={550 1st Avenue},
		city={New York},
		postcode={10016},
		state={NY},
		country={USA}}

	\affiliation[4]{
		organization={Corinne Goldsmith Dickinson Center for Multiple Sclerosis, Department of Neurology, Icahn School of Medicine at Mount Sinai},
		addressline={5 East 98th Street},
		city={New York},
		postcode={10029},
		state={NY},
		country={USA}}

	\affiliation[5]{
		organization={Department of Neurology, New York University School of Medicine},
		addressline={240 E 38th Street},
		city={New York},
		postcode={10016},
		state={NY},
		country={USA}}

	\begin{abstract}
		Since the inception of magnetization transfer (MT) imaging, it has been widely assumed that Henkelman's two spin pools have similar longitudinal relaxation times, which motivated many researchers to constrain them to each other. However, several recent publications reported a $T_1^s$ of the \textit{semi-solid spin pool} that is much shorter than $T_1^f$ of the \textit{free pool}.
		While these studies tailored experiments for robust proofs-of-concept, we here aim to quantify the disentangled relaxation processes on a voxel-by-voxel basis in a clinical imaging setting, i.e., with an effective resolution of 1.24mm isotropic and full brain coverage in 12min.
		To this end, we optimized a \textit{hybrid-state} pulse sequence for mapping the parameters of an unconstrained MT model.
		We scanned four people with relapsing-remitting multiple sclerosis (MS) and four healthy controls with this pulse sequence and estimated $T_1^f \approx 1.84$s and $T_1^s \approx 0.34$s in healthy white matter.
		Our results confirm the reports that $T_1^s \ll T_1^f$ and we argue that this finding identifies MT as an inherent driver of longitudinal relaxation in brain tissue.
		Moreover, we estimated a fractional size of the semi-solid spin pool of $m_0^s \approx 0.212$, which is larger than previously assumed.
		An analysis of $T_1^f$ in normal-appearing white matter revealed statistically significant differences between individuals with MS and controls.
	\end{abstract}

\end{frontmatter}

\footnotetext[1]{corresponding author: Jakob Assl\"ander, Center for Biomedical Imaging, Department of Radiology, New York University School of Medicine, 650 1st Avenue, New York, NY 10016, USA.\\ jakob.asslaender@nyumc.org}

\footnotetext[2]{abbreviations:
	BSA: bovine serum albumin,
	CC: corpus callosum,
	CRB: Cram\'er-Rao bound,
	EDSS: expanded disability status scale,
	FLAIR: fluid-attenuated inversion recovery,
	GM: gray matter,
	MP-RAGE: magnetization-prepared rapid gradient-echo,
	MS: multiple sclerosis,
	MW: myelin water
	NAWM: normal-appearing white matter,
	(q)MT: (quantitative) magnetization transfer,
	ROI: region of interest,
	WM: white matter.
}

\section{Introduction} \label{sec:Introduction}
Longitudinal relaxation is a vital contrast mechanism in magnetic resonance imaging (MRI). For example, the MP-RAGE \citep{Mugler1990} pulse sequence generates excellent gray matter (GM)--white matter (WM) contrast and---compared to mostly $T_2$-weighted pulse sequences like FLAIR \citep{Hajnal.1992}---may be more specific to the underlying tissue changes in multiple sclerosis (MS) lesions \citep{Barkhof.1999,Bagnato2003}.
\citet{Koenig1990} discovered that macromolecules and lipids, in particular myelin, are the source of fast longitudinal relaxation in WM.
Though their experiments were not designed to identify the mechanism through which macromolecules facilitate relaxation, they hypothesized that magnetization transfer (MT) \citep{Wolff1989} is a driving force of relaxation.

Magnetization transfer is commonly described by Henkelman's two-pool model \citep{Henkelman1993}, where one spin pool, the \textit{free} pool, consists of all protons bound in water and is denoted by the superscript $f$, and the other pool, the \textit{semi-solid} pool, consists of protons bound in macromolecules (e.g., proteins and lipids) and is denoted by the superscript $s$.
In standard clinical pulse sequences, one does not directly observe the latter spins because their transversal magnetization relaxes below the noise level before it can be observed ($T_2^s \approx 10\upmu\text{s}$).
However, the exchange of longitudinal magnetization between the two pools alters the free pool's longitudinal magnetization, resulting in bi-exponential relaxation.

The indirect nature of the semi-solid spin pool's impact on the MRI signal entails an entanglement of different parameters.
In order to mitigate the consequent noise amplification, most quantitative MT (qMT) approaches constrain $T_1^s = T_1^f$ resulting in $T_1^s \approx 1.1$s \citep{Yarnykh.2002, Dortch.2011} or, similarly, simply assume that $T_1^s = 1$s \citep{Henkelman1993, Morrison1995a}.
However, recent studies have suggested that $T_1^s \approx 0.3$s and $T_1^f \approx 2$s in white matter at 3T \citep{Helms2009,Gelderen2016,Manning2021,Samsonov2021}.
These studies overcame the challenges of an unconstrained model by either using brain-wide estimates of $T_1^s$ and/or $T_1^f$\citep{Gelderen2016, Samsonov2021} or fitting the MT model to NMR samples \citep{Manning2021} or a single large ROI averaged over multiple participants \citep{Helms2009}.

Our work aims to confirm these findings and to offer evidence in support of Koenig's hypothesis that MT is a key driver of longitudinal relaxation in brain tissue.
Moreover, we aim to provide, for the first time, voxel-wise fits with the unconstrained two-pool MT model.
Key to this advance is a \textit{hybrid state} \citep{Assländer.2019} of the free spin pool that can provide increased efficiency in the encoding and the disentanglement of the MT and relaxation processes \citep{Asslander2020}. Further, we describe the semi-solid spin pool with the \textit{generalized Bloch model} for slight improvements in model accuracy \citep{Assländer.2022u6f}.

We first validated the approach with phantom scans.
Then, we measured reference parameters \textit{in vivo} using 36min scans in participants with multiple sclerosis and healthy controls.
Last, we tested rapid imaging protocols and found that our proposed approach enables unconstrained qMT imaging with an effective resolution of 1.24mm, 1.6mm, and 2.0mm isotropic in 12, 6, and 4 minutes, respectively.


\section{Theory}
\subsection{Magnetization Transfer Model} \label{sec:TheoryMTModel}
\begin{figure}[tb]
	\centering
	\ifOL
		\includegraphics[]{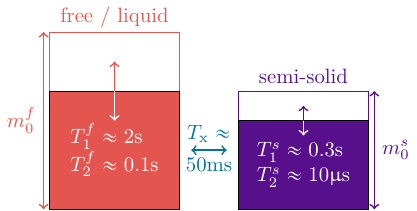}
	\else
		\begin{tikzpicture}[scale = 1, every text node part/.style={align=left}]
			\node [color=Pastrami] at (1.1,3.25) {free / liquid};
			\draw [color=Pastrami] (0,2) rectangle (2.2,3);
			\draw [fill=Pastrami] (0,0) rectangle (2.2,2) node[pos=.5, text = white] {$T_1^f \approx 2$s \\ $T_2^f \approx 0.1$s};
			\draw [->, color = Pastrami, thick] (1.1, 2) -- (1.1, 2.5);
			\draw [->, color = white    , thick] (1.1, 2) -- (1.1, 1.5);

			\draw [<->, color = Pastrami, thick] (-0.1, 0) -- (-0.1, 3) node[left, midway] {$m_0^f$};

			\draw [<->, color = EastRiver, thick] (2.4, 1) -- (3, 1) node[above, midway, text = EastRiver] {$T_{\text{x}} \approx$} node[below, midway, text = EastRiver] {50ms};

			\node [color=NYUpurple] at (4.3,2.25) {semi-solid};
			\draw [color=NYUpurple] (3.2,1.5) rectangle (5.4,2);
			\draw [fill=NYUpurple] (3.2,0) rectangle (5.4,1.5) node[pos=.5, text = white] {$T_1^s \approx 0.3$s \\ $T_2^s \approx 10\upmu$s};
			\draw [->, color = NYUpurple, thick] (4.3, 1.5) -- (4.3, 1.75);
			\draw [->, color = white    , thick] (4.3, 1.5) -- (4.3, 1.25);

			\draw [<->, color = NYUpurple, thick] (5.5, 0) -- (5.5, 2) node[right, midway] {$m_0^s$};

		\end{tikzpicture}
	\fi
	\caption{Sketch of the two-pool magnetization transfer model \citep{Henkelman1993}. This model jointly describes all magnetization arising from protons bound in liquids by the spin pool $m_0^f$, and all magnetization arising from protons bound in macromolecules by the pool $m_0^s$ whose transversal relaxation time is several orders of magnitude shorter. We normalize the thermal equilibrium magnetization to $m_0^f + m_0^s = 1$ and describe the magnetization transfer between the pools by the rate $R_{\text{x}} = 1/T_\text{x}$. The model is governed by Eq.~\eqref{eq:Bloch_McConnell}.
	}
	\label{fig:Model}
\end{figure}
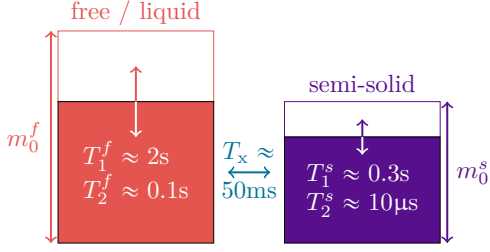

We use the MT model described in \citet{Assländer.2022u6f,Assländer.2024}, which builds on Henkelman's two-pool spin model \citep{Henkelman1993} and captures the two pools with a Bloch-McConnell equation \citep{McConnell1958}:
\begin{widetext}
	\begin{equation}
		\partial_t \begin{pmatrix} x^f \\ y^f \\ z^f \\ x^s \\ z^s \\ 1 \end{pmatrix} = \begin{pmatrix}
			-R_2^f    & -\omega_z & \omega_y                    & 0                                    & 0                           & 0           \\
			\omega_z  & -R_2^f    & 0                           & 0                                    & 0                           & 0           \\
			-\omega_y & 0         & -R_1^f - R_{\text{x}} m_0^s & 0                                    & R_{\text{x}} m_0^f          & m_0^f R_1^f \\
			0         & 0         & 0                           & -R_2^{s,l}(R_2^s,\alpha,T_\text{RF}) & \omega_y                    & 0           \\
			0         & 0         & R_{\text{x}} m_0^s          & -\omega_y                            & -R_1^s - R_{\text{x}} m_0^f & m_0^s R_1^s \\
			0         & 0         & 0                           & 0                                    & 0                           & 0
		\end{pmatrix} \begin{pmatrix} x^f \\ y^f \\ z^f \\ x^s \\ z^s \\ 1 \end{pmatrix}.
		\label{eq:Bloch_McConnell}
	\end{equation}
\end{widetext}
The \textit{free} pool, sketched in red in Fig.~\ref{fig:Model}, captures all protons bound in liquids where fast molecular motion causes an exponential relaxation of the transversal magnetization with a characteristic $T_2^f \gtrsim 50$ms \citep{Bloembergen1948}.
The free pool's magnetization is described by the Cartesian coordinates $x^f$, $y^f$, $z^f$, the off-resonance frequency is described by $\omega_z$, and the Rabi frequency of the RF pulses by $\omega_y$. For readability, we here use relaxation rates ($R_{1,2}^{f,s} = 1 / T_{1,2}^{f,s}$).
The magnetization components $x^s,z^s$ of the \textit{semi-solid} spin pool, sketched in purple in Fig.~\ref{fig:Model}, capture all protons bound in large molecules such as lipids. The motion of such molecules is restricted, resulting in a much faster and non-exponential relaxation with a characteristic time constant of $T_2^s \approx 10\upmu$s, which prevents a direct detection of this pool with standard clinical MRI.
Within the brain parenchyma, we assume the decay characteristics associated with a super-Lorentzian lineshape \citep{Morrison1995a}.
The non-exponential characteristics of this lineshape prohibit a description with the original Bloch equations, but such dynamics can be described with the \textit{generalized Bloch model} \citep{Assländer.2022u6f}.
For computational efficiency, we can approximate the non-exponential decay by an effective exponential decay with a \textit{linearized} relaxation rate $R_2^{s,l}(R_2^s,\alpha,T_\text{RF})$. While exponential and non-exponential decays necessarily deviate, we can identify an $R_2^{s,l}$ that results in the same magnetization at the end of an RF pulse. To this end, $R_2^{s,l}$ depends on the flip angle $\alpha$ and the duration $T_\text{RF}$ of respective RF-pulse in addition to the biophysical parameter $R_2^s$.
We neglect the $y^s$ component assuming, without loss of generality, $\omega_x=0$ and given that $R_2^{s,l} \gg \omega_z$.
The exchange rate $R_{\text{x}}$ captures exchange processes between the pools.
A sixth dimension is added to allow for a compact notation of the longitudinal relaxation to a non-zero thermal equilibrium.

Throughout the literature, multiple normalizations of $m_0^s$ have been used. Here, we use $m_0^f + m_0^s = 1$ so that $m_0^s$ describes the fraction of the overall spin pool, a definition which has also been used, e.g., by \citet{Yarnykh.2012}.
Other papers, such as \citet{Henkelman1993,Gochberg.2003}, measure $\tilde{m}_0^s = m_0^s / m_0^f$ or, equivalently, normalize to $\tilde{m}_0^f = 1$.
The conversion between the two definitions is simply
$m_0^s = \tilde{m}_0^s / (1 + \tilde{m}_0^s)$ and
$\tilde{m}_0^s = m_0^s / (1 - m_0^s)$.

\subsubsection{Comparison to constrained MT models} \label{sec:TheoryBiexpModel}
In the absence of RF pulses ($\omega_y = 0$), we can isolate the longitudinal components of Eq. \eqref{eq:Bloch_McConnell}:
\begin{equation}
	\partial_t \begin{pmatrix} z^f \\ z^s \\ 1 \end{pmatrix} = \begin{pmatrix}
		-R_1^f - R_\text{x} m_0^s & R_\text{x} m_0^f          & m_0^f R_1^f \\
		R_\text{x} m_0^s          & -R_1^s - R_\text{x} m_0^f & m_0^s R_1^s \\
		0                         & 0                         & 0
	\end{pmatrix} \begin{pmatrix} z^f \\ z^s \\ 1 \end{pmatrix}.
	\label{eq:Longitudinal}
\end{equation}
An eigendecomposition of the Hamiltonian in Eq.~\eqref{eq:Longitudinal} has three distinct eigenvalues \citep{Henkelman1993, Gochberg.2003,Yarnykh.2012}.
One is zero and corresponds to thermal equilibrium.
The smaller remaining eigenvalue (in the absolute value) can be considered an apparent relaxation rate of the free pool $R_1^{f,a}$ that is approximated by the following Taylor expansion at $R_1^s = R_1^f$:
\begin{equation}
	R_1^{f,a} \approx R_1^f + m_0^s (R_1^s - R_1^f) - \frac{m_0^f m_0^s (R_1^s - R_1^f)^2}{R_\text{x}} .
	\label{eq:R1a}
\end{equation}
The MT contributions to $R_1^{f,a}$ therefore depend foremost on the macromolecular pool size $m_0^s$ and the two relaxation rates. Higher order terms additionally depend on the exchange rate $R_\text{x}$.
Eq.~\eqref{eq:R1a} shows that $R_1^{f,a} \approx R_1^f$ is a reasonable approximation only if $m_0^s(R_1^s - R_1^f) \ll R_1^f$.
Otherwise, this linear correction term contributes significantly to $R_1^{f,a}$, making MT an important driver of longitudinal relaxation.
For example, let us assume $R_1^f = 0.5/$s, $R_1^s = 3$/s, and $m_0^s = 0.2$. In this case, the linear correction term is 0.5/s and, thus, $R_1^{f,a} \approx 1.0/\text{s} \not\approx R_1^f$.

The largest eigenvalue (in absolute value) is given by
\begin{equation}
	R_\text{x}^a \approx (R_\text{x} + R_1^f) + m_0^f (R_1^s - R_1^f) + \frac{m_0^f m_0^s (R_1^s - R_1^f)^2}{R_\text{x}},
	\label{eq:R1cross}
\end{equation}
which is dominated by the exchange rate $R_\text{x}$ for many tissues. Hence, it can be interpreted as a cross-relaxation term and we henceforth refer to it as the apparent exchange rate.

From the eigenvectors, we can derive a Taylor expansion of the apparent semi-solid pool size (\ref{sec:AppendixEigendecomposition}):
\begin{equation}
	m_0^{s,a} \approx m_0^s \left(1 - \frac{2 m_0^f (R_1^s - R_1^f)}{R_\text{x}}\right).
	\label{eq:m0s_Taylor}
\end{equation}
Eq.~\eqref{eq:m0s_Taylor} reveals that $m_0^s$ is underestimated when assuming $R_1^s = R_1^f$.
To give a sense of the magnitude of this bias, we can insert the above example values and further assume $R_\text{x} = 15/$s, which results in $m_0^{s,a} \approx 0.15$ instead of the underlying $m_0^s = 0.2$.

\section{Methods}
\subsection{Pulse sequence design}
As mentioned above, we utilize the hybrid state \citep{Assländer.2019} and its flexibility to encode and disentangle the different relaxation mechanisms.
Similar to balanced SSFP sequences \citep{Carr1958}, we balance all gradient moments in each $T_\text{R}$.
On the other hand, we vary the flip angle and the duration of the RF pulses.
During slow flip angle variations, the direction of the magnetization establishes a steady state and adiabatically transitions between the steady states associated with different flip angles.
As we showed in \citet{Assländer.2019}, moderate change rates of the flip angle simultaneously yield a transient state of the magnetization's magnitude, and we call this combination the \textit{hybrid state}.
It combines the tractable off-resonance characteristics of the bSSFP sequence, particularly the refocusing of intra-voxel dephasing \citep{Carr1958,Scheffler2003}, with the encoding potential of the transient state.

Our pulse sequence consists of a rectangular $\pi$ inversion pulse, flanked by crusher gradients, followed by a train of rectangular RF pulses with varying flip angles and pulse durations.
The RF phase is incremented by $\pi$ between consecutive RF pulses.
The pulses are separated by a $T_\text{R} = 3.5\text{ms}$, which is approximately the minimal $T_\text{R}$ with which we can perform gradient encoding with $|k_{\max}| = \pi/1$mm and avoid stimulating the peripheral nerves.
After 1142 RF-pulses, i.e., after a \textit{cycle time} of 4s, the remaining magnetization is inverted by the subsequent $\pi$ pulse, then the same pulse train is repeated.

The relaxation and MT processes are encoded with two established mechanisms.
First, the $T_2$-selective inversion pulse inverts the free pool while keeping the semi-solid pool largely unaffected. As described by \citet{Gochberg.2003}, this induces a bi-exponential inversion recovery curve of the free pool composed of its intrinsic longitudinal relaxation and cross-relaxation to the semi-solid spin pool.
Second, we can use the flip angle and the pulse duration to control the different relaxation paths.
In good approximation, the RF-pulse duration only affects the saturation of the semi-solid spin pool's longitudinal magnetization \citep{Gloor2008}.
In contrast, changes in the flip angle affect the relaxation processes of the free pool \citep{Asslander2019b,Assländer.2019}, the magnetization transfer between the two pools, and the saturation of the semi-solid spin pool \citep{Gloor2008}.
More details on this interplay can be found in \citet{Assländer.2024}.

\subsection{Numerical optimization of the pulse train} \label{sec:MethodsNumericalOptimzation}
We numerically optimized the flip angles and pulse durations of RF-pulse trains based on these two encoding mechanisms.
The optimization objective was the Cram\'er-Rao bound (CRB) \citep{Rao1945,Cramer1946}, which predicts the noise variance of a fully efficient unbiased estimator. We note that least squares fitting and the neural network-based fitting used in this article (cf. Section \ref{sec:nn}) are, strictly speaking, neither fully efficient nor unbiased \citep{Newey.1994,Wu.1981}.
Nonetheless, the CRB can be used as a proxy for the ``SNR-efficiency'' or ``conditioning'' \citep{Zhao2019, Haldar.2019} and we adopt this heuristic here.

We calculated the CRB as described in \citet{Assländer.2024} and optimized for the CRBs of the relaxation rates and the other model parameters.
We optimized a separate pulse train for each of the biophysical parameters $m_0^s$, $R_1^f$, $R_2^f$, $R_\text{x}$, $R_1^s$, and $T_2^s$, while additionally accounting for $\omega_z$, $B_1^+ = \omega_y / \omega_y^{\text{nominal}}$, and the scaling factor $M_0$ as unknowns, where $M_0$ jointly describes the overall spin density and receive-coil sensitivity profiles.
Additionally, we optimized a pulse train for the sum of the CRBs of all biophysical parameters, normalized with respective squared parameter values to resemble the inverse squared SNR.
We performed all simulations and CRB calculations with
$m_0^s = 0.25$,
$R_1^f = 0.5/\text{s}$,
$R_2^f = 15.4/\text{s}$,
$R_\text{x} = 20/\text{s}$,
$R_1^s = 2$/s,
$T_2^s = 10\upmu\text{s}$, $\omega_z = 0$, and $B_1^+ = 1$.
The resulting spin trajectories (Fig.~\ref{fig:Spin_Dynamics}) and the corresponding CRB values (Tab.~\ref{tab:CRB}) are discussed in the supporting information Section~\ref{supsec:NumericalOptimizations}.
Supporting Section~\ref{supSec:Noise} and Figs.~\ref{fig:Phantom_PNRCG}, \ref{fig:Phantom_repLLRvCG}, and \ref{fig:InVivo_CRB} connect the CRB to experimental noise measurements.

\subsection{Phantom scan} \label{sec:MethodsPhatom}
We built a custom phantom composed of cylindrical 50mL tubes filled with different concentrations of thermally cross-linked bovine serum albumin (BSA).
We mixed the BSA powder (5\%, 10\%, \ldots, 35\% of the total weight) with distilled water and stirred it at 30$^\circ$C until the BSA was fully dissolved.
We filled 7 tubes with the resulting solutions and thermally cross-linked them in a water bath at approximately 90$^\circ$C for 10 minutes.

We scanned this phantom on a 1.5T Sola and  2.9T Prisma scanner (Siemens, Erlangen, Germany).
We performed a 6min scan with each of 6 individual optimizations, resulting in a 36min overall scan time. For each 6min scan, the RF pattern is repeated 90 times, during which we acquire 3D radial k-space spokes with nominal 1.0mm isotropic resolution (defined by $|k_{\max}| = \pi/1$mm).
The sampled k-space covers the insphere of the typically acquired 1.0mm k-space cube. By comparing the covered k-space volume, we estimate an effective resolution of 1.24mm, which we report throughout this paper \citep{Pipe.2011}.
We note that the stated effective resolution does not account for blurring introduced by undersampling in combination with a regulized reconstruction.
We changed the direction of the k-space spokes with a 2D golden means pattern \citep{Winkelmann2007,Chan.2009} that is reshuffled to improve the k-space coverage for each time point and to minimize eddy current artifacts \citep{Flassbeck.2023ilg}.

\subsection{In vivo scans}
Each participant's informed consent was obtained before the scan following a protocol that was approved by the NYU School of Medicine Institutional Review Board.
To establish high-quality reference data, we performed in vivo scans of 4 individuals with clinically established relapsing-remitting MS (extended disability status scale (EDSS) 1.0--2.5, unknown for one participant; no recent history of relapses; age $37.5 \pm 8.7$; 3 female) and 4 healthy controls (age $28.8 \pm 5.6$; 3 female) with a 2.9T Prisma scanner and the 36min protocol described in Section \ref{sec:MethodsPhatom}.
In addition to the hybrid-state scans, we acquired 3D MP-RAGE and FLAIR scans, each with a 1.0mm isotropic resolution.

To test more clinically feasible scan times, we scanned an additional participant with MS with three ``rapid" protocols with different effective resolutions:
\begin{itemize}
	\setlength{\parskip}{0pt}
	\setlength{\itemsep}{0pt plus 1pt}
	\item 1.24mm isotropic in 12min
	\item 1.6mm isotropic in 6min
	\item 2.0mm isotropic in 4min.
\end{itemize}

\subsection{Image reconstruction} \label{sec:MethodsImageRecon}
We performed retrospective motion correction similar to \citet{Kurzawski.2020}.
However, our approach deviates from Kurzawski et al. in one key aspect: Instead of using an SVD to maximize the first coefficient's signal intensity, we utilize a generalized eigendecomposition \citep{Kim.2021} to maximize the contrast between brain parenchyma and CSF \citep{Flassbeck.2024}.
We reconstructed images directly in the space spanned by three basis functions associated with the generalized eigendecomposition \citep{Tamir2017} and used a total variation penalty along time to reduce undersampling artifacts \citep{Feng2014}.
The reconstructions were performed with a spatial resolution of 4mm isotropic and a temporal resolution of 4s. The images corresponding to the first coefficient were co-registered using SPM12 and the extracted transformation matrices were subsequently applied to the k-space data (translations) and trajectory (rotations) to correct the full-resolution reconstruction.

We reconstructed the images with sub-space modeling \citep{Liang.2007, Huang.2012, Christodoulou.2018, McGivney.2014, Tamir2017, Asslander2018, Zhao2018}, i.e., we reconstructed coefficient images in the sub-space spanned by singular vectors from a coarse dictionary of signals (or fingerprints) \citep{McGivney.2014,Tamir2017,Asslander2018} and their orthogonalized gradients \citep{Mao.2023yo}.
We used the optISTA algorithm \citep{Jang.2023}, incorporating sensitivity encoding \citep{Sodickson1997, Pruessmann.2001} and locally low-rank regularization \citep{Lustig2007, Trzasko2011, Zhang2015} to reduce residual undersampling artifacts and noise.
We implemented this reconstruction in \textit{Julia} and made the source code publicly available (cf. \ref{sec:data_availability}).
A more detailed description of the reconstruction can be found in \citet{Tamir2017} and \citet{Asslander2018}.

For the 36min in vivo scans, we used separate sub-spaces for each 6min sub-scan, implemented as a block-diagonal matrix to permit joint regularization.
For the phantom scan and the rapid protocols, we reconstructed all data of the 6 sub-scans into a joint 15-dimensional subspace with otherwise identical settings.

\subsection{Model fitting} \label{sec:nn}
For computational efficiency and robustness, we used neural networks to fit the MT model, voxel by voxel, to the reconstructed coefficient images \citep{Cohen2018,Nataraj.2018,Duchemin2020,Zhang.2022,Mao.2023jtc}.
This approach includes a data-driven $B_0$ and $B_1^+$ correction as detailed in \citet{Assländer.2024}.
For each voxel, the complex-valued coefficients are normalized by the first coefficient and split into real and imaginary parts, which are the inputs to the network.
The network retains a similar overall architecture to the design described in Fig. 2 of \citet{Zhang.2022}:
11 fully connected layers with skip connections and batch normalization, and a maximum layer width of 1024.
The network estimates all 6 biophysical parameters of the unconstrained MT model. For both reconstruction protocols, we trained networks using the Rectified ADAM optimizer \citep{Liu.2019} to convergence with individually-tuned learning rates.
For more details, we refer to \citet{Zhang.2022} and \citet{Mao.2023jtc}.

\subsection{Region of interest analysis}
For the 36min reference scans, we registered the skull-stripped \citep{Hoopes.2022} qMT maps and the FLAIR images to the MP-RAGE  with the FreeSurfer package (``mri\_robust\_register") \citep{Reuter.2010}. We also used FreeSurfer (``recon-all") to segment the brain based on the MP-RAGE and the FLAIR \citep{Fischl.2002,Fischl.2004}.
We extracted region of interest (ROI) masks for the entire normal-appearing white matter (NAWM), several WM subregions, the cortical GM, and subcortical GM structures.
To ensure that MS lesions were excluded from the ROIs, we calculated lesion masks with an in-house developed deep learning model based on the nnUNet framework \citep{Isensee.2021} using the FLAIR images.
The automated lesion segmentations were manually adjusted by FLR and ESB and subtracted from the ROI masks.
After that, we eroded the outmost layer of voxels of each ROI to reduce partial volume effects with other tissues and to ensure that all ROI voxels are at least one voxel away from any lesion.

\begin{figure}[htbp]
	\centering
	\ifOL
		\includegraphics[]{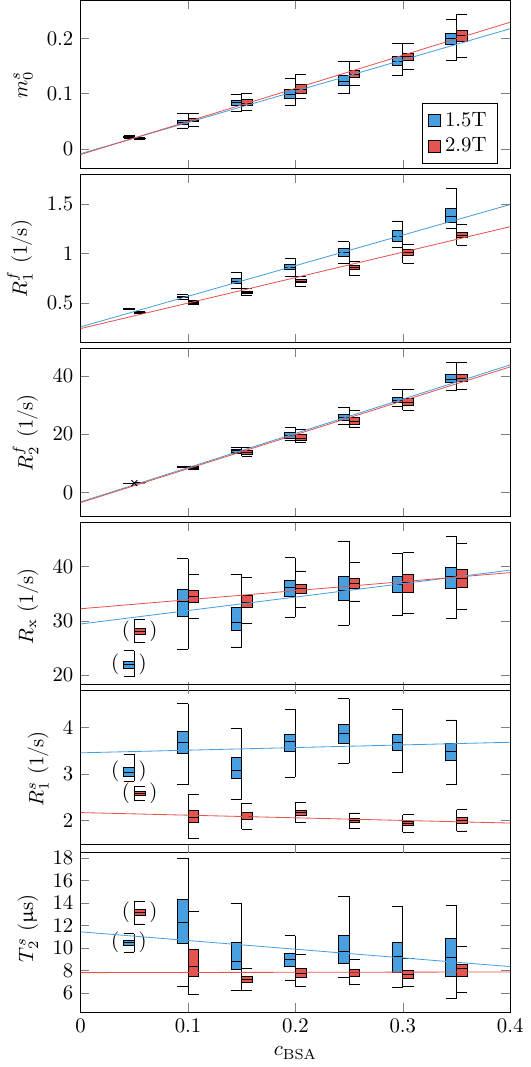}
	\else
		\input{Figures/Phantom_linearRegression.tex}
	\fi
	\caption{Phantom validation. Seven tubes filled with different concentrations of bovine serum albumin (BSA) were imaged at 1.5T and 2.9T. The box plots represent the median, 1\textsuperscript{st} and 3\textsuperscript{rd} quartile, and the whiskers the 1.5x the inter-quartile range or the maximum range, whichever is smaller. The median values of each tube's qMT estimates were fitted with a general linear model with the BSA concentration ($c_\text{BSA}$) and the field strength ($B_0$) as independent variables. The fitted coefficients are listed in Tab.~\ref{tab:Phantom}. The brackets indicate outliers that were excluded from the GLM regression due to unstable parameter estimation of the semi-solid pool's characteristic, likely caused by the small pool size.}
	\label{fig:Phantom}
\end{figure}

\begin{table}[tb]
	\centering
	\newcommand{\g}{\cellcolor[gray]{0.8}}
	\begin{tabular}{l|c|c|c|c||c}
		                                   & $a_0$    & $a_\text{BSA}$ & $a_{B_0} (\sfrac{1}{\text{T}})$ & $a_2 (\sfrac{1}{\text{T}})$ & $R^2$ \\ \hline
		$m_0^s$                            & \g-0.007 & 0.53           & \g-0.0011                       & \g0.024                     & 0.99  \\
		$R_1^f (\sfrac{1}{\text{s}})$      & 0.28     & 3.6            & \g-0.012                        & -0.37                       & 0.99  \\
		$R_2^f (\sfrac{1}{\text{s}})$      & \g-2.9   & 119            & \g-0.16                         & \g-0.81                     & 0.99  \\
		$R_\text{x} (\sfrac{1}{\text{s}})$ & 26       & \g34           & \g2.0                           & \g-5.9                      & 0.68  \\
		$R_1^s (\sfrac{1}{\text{s}})$      & 4.8      & \g1.8          & -0.92                           & \g-0.8                      & 0.95  \\
		$T_2^s (\upmu \text{s})$           & 15       & \g-1.6         & -2.6                            & \g5.7                       & 0.67  \\
	\end{tabular}
	\caption{Coefficients of a generalized linear model fit of the phantom data shown in Fig.~\ref{fig:Phantom}. The fitted function has the form $a_0 + a_\text{BSA} \cdot c_\text{BSA} + a_{B_0} \cdot B_0 + a_2 \cdot B_0 \cdot c_\text{BSA}$.
		The white background identifies coefficients that differ from zero at a 95\% confidence interval.
		The gray background identifies coefficients for which such determination cannot be made.
		The rightmost column denotes the coefficient of determination.
	}
	\label{tab:Phantom}
\end{table}

\begin{figure*}[tb]
	\centering
	\ifOL
		\includegraphics[]{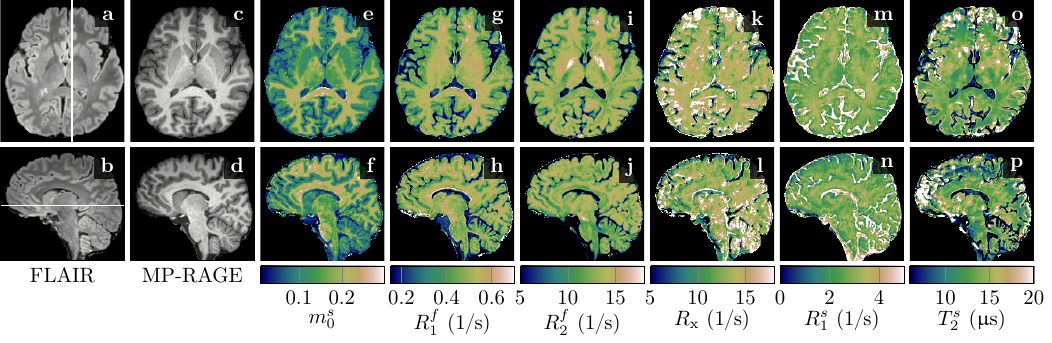}
	\else
		\input{Figures/In_Vivo_Maps_Control.tex}
	\fi
	\caption{
	Comparison of clinical contrasts (\textbf{a}-\textbf{d}) and quantitative magnetization transfer (qMT) maps (\textbf{e}-\textbf{p}) in a healthy volunteer. The qMT maps have an effective resolution of 1.24mm isotropic (acquired in 36min) compared to the 1mm isotropic of the clinical contrasts. We display here relaxation rates ($R_{1,2}^{f,s} = 1/T_{1,2}^{f,s}$), where the superscripts $f$ and $s$ indicate the \textit{free} and \textit{semi-solid} pools, respectively.
	The size of the semi-solid spin pool is normalized by $m_0^s + m_0^f = 1$, and $R_\text{x}$ denotes the exchange rate between the two pools.
	}
	\label{fig:InVivo_ctrl}
\end{figure*}

\section{Results}
\subsection{Phantom scan} \label{sec:ResultsPhantom}
The phantom validation aims to identify the parameters' dependency on the sample's protein (BSA) concentration and the magnetic field strength and to compare these findings to previous work and our general understanding of relaxation.
To this end, we performed a generalized linear model (GLM) fit of the data with the BSA concentration ($c_\text{BSA}$) and the magnetic field strength ($B_0$) as independent variables (Fig.~\ref{fig:Phantom} and Tab.~\ref{tab:Phantom}).

The estimates of the semi-solid spin pool size are consistent with the linear model $m_0^s = a_\text{BSA} \cdot c_\text{BSA}$, i.e., we can reject neither the null hypothesis that $m_0^s$ is independent of $B_0$ nor that it vanishes at $c_\text{BSA}=0$.

For $R_1^f$, our data supports the linear model $R_1^f = a_0 + a_\text{BSA} \cdot c_\text{BSA} + a_2  \cdot c_\text{BSA} \cdot B_0$. I.e., the data suggests a linear dependence of $R_1^f$ on the BSA concentration where the slope also depends on the field strength.
By contrast, we did not observe a dependency of $R_1^f$ of pure water ($c_\text{BSA}=0$) on $B_0$.
This finding is consistent with the very small change predicted by the Bloembergen-Purcell-Pound (BPP) theory (approximately 0.004\% for pure water with a correlation time $\tau_c = 5 \text{ps}$; \citep{Bloembergen1948}).
Further, the intercept $a_0 = 0.28/\text{s}$ with a 95\% confidence interval $[0.15, 0.41]/\text{s}$ agrees with the $0.255/\text{s}$ predicted by the BPP theory for 1.5T and 2.89T (the BPP theory predicts an identical rate at both field strengths within the indicated precision).

For $R_2^f$, our data is consistent with a linear dependence on the BSA concentration and no dependence on $B_0$ ($R_2^f = a_\text{BSA} \cdot c_\text{BSA}$).
The latter is also consistent with the very small change predicted by the BPP theory (approximately 0.001\%).
While the negative intercept $a_0 = -2.9/\text{s}$ is not physical, the 95\% confidence interval $[-7.6, 1.8]/\text{s}$ includes the $R_2^f \approx 0.255/\text{s}$ predicted by the BPP theory.

We detected no dependence of $R_1^s$ on $c_\text{BSA}$, but a statistically significant dependence on $B_0$ ($R_1^s = a_0 + a_{B_0} \cdot B_0$), which is consistent with the reports of \citet{Wang.2020}. For $R_\text{x}$, we observe neither a $c_\text{BSA}$ nor a $B_0$ dependency, and for $T_2^s$, the $B_0$ dependency is just above the 95\% significance threshold.

Beyond the linear model, we observe increased variability of $R_\text{x}$, $R_1^s$, and $T_2^s$ estimates with decreasing $c_\text{BSA}$, which likely stems from the smaller spin pool size. For this reason, we excluded the most extreme case ($c_\text{BSA} = 0.05$) from the GLM fits as indicated by the brackets in Fig.~\ref{fig:Phantom}.


\begin{table*}[tb]
	\centering
	\begin{tabular}{l|c|c|c|c|c|c}
		             & $m_0^s$           & $T_1^f$ (s)       & $T_2^f$ (ms)   & $R_\text{x}$ (1/s) & $T_1^s$ (s)       & $T_2^s$ ($\upmu$s) \\ \hline
		entire WM    & $0.212 \pm 0.022$ & $1.84 \pm 0.17$   & $76.9 \pm 8.3$ & $13.6 \pm 1.1$     & $0.34 \pm 0.10$   & $12.5 \pm 1.8$     \\
		anterior CC  & $0.237 \pm 0.032$ & $1.77 \pm 0.26$   & $69.9 \pm 6.5$ & $13.4 \pm 1.7$     & $0.349 \pm 0.045$ & $14.5 \pm 2.7$     \\
		posterior CC & $0.235 \pm 0.038$ & $1.80 \pm 0.17$   & $76.3 \pm 5.6$ & $13.5 \pm 1.9$     & $0.350 \pm 0.049$ & $12.6 \pm 1.2$     \\
		cortical GM  & $0.098 \pm 0.026$ & $2.46 \pm 0.56$   & $83 \pm 15$    & $14.0 \pm 3.1$     & $0.42 \pm 0.40$   & $14.4 \pm 3.9$     \\
		Caudate      & $0.113 \pm 0.020$ & $1.95 \pm 0.16$   & $73.3 \pm 4.4$ & $13.8 \pm 2.2$     & $0.432 \pm 0.095$ & $15.1 \pm 2.3$     \\
		Putamen      & $0.118 \pm 0.018$ & $1.84 \pm 0.14$   & $67.4 \pm 5.0$ & $14.9 \pm 1.8$     & $0.385 \pm 0.048$ & $15.4 \pm 2.2$     \\
		Pallidum     & $0.164 \pm 0.025$ & $1.664 \pm 0.088$ & $59.3 \pm 5.6$ & $15.8 \pm 1.8$     & $0.351 \pm 0.038$ & $14.9 \pm 2.4$     \\
		Thalamus     & $0.158 \pm 0.029$ & $2.02 \pm 0.27$   & $70.8 \pm 6.2$ & $14.2 \pm 1.9$     & $0.396 \pm 0.061$ & $13.0 \pm 1.8$     \\
		Hippocampus  & $0.097 \pm 0.024$ & $2.65 \pm 0.84$   & $91 \pm 15$    & $15.3 \pm 2.7$     & $0.376 \pm 0.098$ & $13.0 \pm 3.2$     \\
	\end{tabular}
	\caption{
		Region of interest (ROI) analysis in healthy controls. The ROIs were determined by segmenting the co-registered MP-RAGE images with the \textit{FreeSurfer} software. The values represent the mean and standard deviation of all voxels from 4 healthy participants. WM is short for white matter and CC for corpus callosum.
	}
	\label{tab:ROImean}
\end{table*}

\subsection{In vivo reference scans} \label{sec:ResultsControl}
Fig. \ref{fig:InVivo_ctrl} demonstrates the feasibility of unconstrained qMT imaging with a hybrid-state pulse sequence, i.e., encoding all 6 biophysical parameters on a voxel-by-voxel basis.
By comparing the qMT maps to the routine clinical contrasts, we observe overall good image quality in $m_0^s$, $R_1^f$, and $R_2^f$.
However, the cerebellum reveals a slightly reduced effective resolution compared to the nominally equivalent resolution of the MP-RAGE (Fig. \ref{fig:InVivo_ctrl}d vs. f,h,\dots).
The $R_\text{x}$ and $R_1^s$ maps exhibit reduced image quality, consistent with their higher CRB values (Tab. \ref{tab:CRB}).
Also consistent with its large CRB, the $T_2^s$ map has the highest noise and artifact levels, which might also be, in part, due to a residual $B_1^+$ artifact caused by incomplete spoiling of the inversion pulse.
We also find subtle residual $B_1^+$ artifacts in $R_2^f$ (Fig.~\ref{fig:InVivo_ctrl}i,j, and Fig. \ref{fig:InVivo_MS}c) and residual $B_0$ artifacts in a few voxels at the center of the bSSFP banding artifact (Fig. \ref{fig:InVivo_ctrl}f,h,\dots \; at the base of the frontal cortex). Overall, however, we observe good performance of the data-driven $B_0$ and $B_1^+$ correction.

Among all qMT parameters, we observe the largest quantitative GM-WM contrast in $m_0^s$, followed by $R_1^f$.
In $R_2^f$, however, we observe only a subtle contrast between cortical GM and WM. An ROI analysis confirms this finding:
we estimated $T_2^f = (83 \pm 15)$ms and $(76.9 \pm 8.3)$ms for cortical GM and WM, respectively, which is a smaller difference compared to the difference between previously reported values ($(99 \pm 7)$  vs. $(69 \pm 3)$ms; \citep{Stanisz.2005}).
Consistent with previous reports, we observe the shortest $T_2^f = (59.3 \pm 5.6)$ms in the pallidum (Fig. \ref{fig:InVivo_ctrl}i).
The exchange rate $R_\text{x}$, $R_1^s$, and $T_2^s$ exhibit little GM-WM contrast and we note that the most prominent contrast in $R_\text{x}$ and $R_1^s$ occurs in voxels subject to partial volume effects and in CSF.
In voxels with partial volume, the model might be inaccurate and the small $m_0^s$ makes estimates of semi-solid spin-pool characteristics unreliable, in particular with an unconstrained model, as previously demonstrated by \citet{Dortch.2018}.
Estimates of the unconstrained MT model's parameters are reported in Tab.~\ref{tab:ROImean} for selected WM and GM structures.

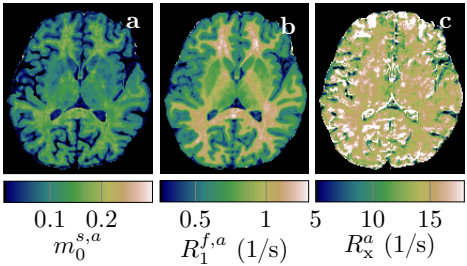
\begin{figure}[tb]
	\centering
	\ifOL
		\includegraphics[]{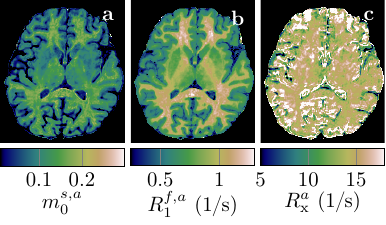}
	\else
		\begin{tikzpicture}[scale = 1, every text node part/.style={align=left}]
    \begin{axis}[%
            width={2.1cm},
            height={2.1cm*1.143},
            axis on top,
            scale only axis,
            xmin=0,
            xmax=1,
            ymin=0,
            ymax=1,
            xtick = \empty,
            ytick = \empty,
            name=m0s,
            colormap name = gist_earth,
            colorbar horizontal,
            point meta min=0.01,
            point meta max=0.299,
            colorbar style={xlabel=$m_0^{s,a}$, height=0.3cm, yshift=0.2cm, xlabel style = {yshift = 0.15cm}, xticklabel style={/pgf/number format/fixed, /pgf/number format/precision=2}},
        ]
        \addplot graphics [xmin=0,xmax=1,ymin=0,ymax=1] {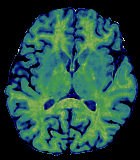};
        \node[text=white, anchor = north east] at (rel axis cs:  0.975,0.975) {\textbf{a}};
    \end{axis}

    \begin{axis}[%
            width={2.1cm},
            height={2.1cm*1.143},
            axis on top,
            scale only axis,
            xmin=0,
            xmax=1,
            ymin=0,
            ymax=1,
            hide axis,
            name=R1f,
            at=(m0s.right of north east),
            anchor=north west,
            xshift = 0.1cm,
            colormap name = gist_earth,
            colorbar horizontal,
            point meta min=0.25,
            point meta max=1.3,
            colorbar style={xlabel=$R_1^{f,a}~(1/\text{s})$, height=0.3cm, yshift=0.2cm, xlabel style = {yshift = 0.15cm}, xticklabel style={/pgf/number format/fixed, /pgf/number format/precision=2}},
        ]
        \addplot graphics [xmin=0,xmax=1,ymin=0,ymax=1] {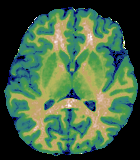};
        \node[text=white, anchor = north east] at (rel axis cs:  0.975,0.975) {\textbf{b}};
    \end{axis}

    \begin{axis}[%
            width={2.1cm},
            height={2.1cm*1.143},
            axis on top,
            scale only axis,
            xmin=0,
            xmax=1,
            ymin=0,
            ymax=1,
            hide axis,
            name=R2f,
            at=(R1f.right of north east),
            anchor=north west,
            xshift = 0.1cm,
            align=center,
            colormap name = gist_earth,
            colorbar horizontal,
            point meta min=5,
            point meta max=18,
            colorbar style={xlabel=$R_\text{x}^a~(1/\text{s})$, height=0.3cm, yshift=0.2cm, xlabel style = {yshift = 0.15cm}, xticklabel style={/pgf/number format/fixed, /pgf/number format/precision=2}},
        ]
        \addplot graphics [xmin=0,xmax=1,ymin=0,ymax=1] {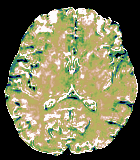};
        \node[text=white, anchor = north east] at (rel axis cs:  0.975,0.975) {\textbf{c}};
    \end{axis}
\end{tikzpicture}
	\fi
	\caption{Apparent quantitative MT maps when assuming $T_1^s = T_1^f$ in a healthy volunteer. The maps were calculated voxel-wise with Eqs.~\eqref{eq:R1a}--\eqref{eq:m0s_Taylor} and based on the maps depicted in Fig.~\ref{fig:InVivo_ctrl}. Note the different color scale in $R_1^{f,a}$ compared to Fig.~\ref{fig:InVivo_ctrl}.}
	\label{fig:InVivo_ctrl_R1a}
\end{figure}

\begin{figure}[h!]
	\centering
	\ifOL
		\includegraphics[]{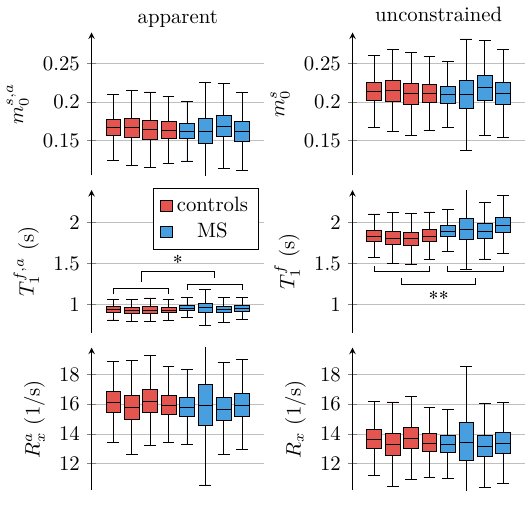}
	\else
		\input{Figures/ROI_unconstraint_vs_Taylor.tex}
	\fi
	\caption{Comparison of apparent qMT parameter estimates when assuming $T_1^s = T_1^f$ (Eqs.~\eqref{eq:R1a}--\eqref{eq:m0s_Taylor}) to unconstrained parameter estimates. The box plots pool all normal-appearing white matter voxels of each participant.
		The markers $*$ and $**$ indicate statistically significant differences at the $p < 0.05$ and $p < 0.01$ levels in a comparison of each subject's median qMT parameter estimates between participants with MS and controls.
	}
	\label{fig:ROI_apparent}
\end{figure}

\subsubsection{Comparison to constrained MT models}
Fig.~\ref{fig:InVivo_ctrl_R1a} depicts the apparent qMT parameters associated with a $T_1^s = T_1^f$ constrained model (Eqs.~\eqref{eq:R1a}--\eqref{eq:m0s_Taylor}). Fig.~\ref{fig:ROI_apparent} compares the apparent qMT parameters of those fitted with the unconstrained model, with more brain ROIs analyzed in Supporting Tab.~\ref{tab:ROImean_apparent}. Below, we discuss the salient differences in white and cortical gray matter.

\paragraph{White matter}
With the unconstrained model, we estimated a substantially different $T_1^f = (1.84 \pm 0.17)$s from a $T_1^s = (0.34 \pm 0.10)$s. Using Eq.~\eqref{eq:R1a} to calculate the apparent $T_1^{f,a}$, we estimate $T_1^{f,a} = (0.941 \pm 0.069)\text{s}$, which approximately matches literature values (($1.084 \pm 0.045$)s \citep{Stanisz.2005}).

With the unconstrained MT model, we estimated $m_0^s = 0.212 \pm 0.022$, consistent with literature estimates using the same model ($0.172 \pm 0.043$; \citep{Helms2009}).
The apparent pool size (Eq.~\eqref{eq:m0s_Taylor}) is $m_0^{s,a} = 0.151 \pm 0.022$, which also matches the constrained estimates in the literature ($0.139 \pm 0.028$; \citep{Stanisz.2005} and $0.118 \pm 0.050$; \citep{Helms2009}).

The exchange rate estimated with the unconstrained MT model, $R_\text{x} = (13.6 \pm 1.1)$/s, is slightly lower compared to the corresponding literature (($18.1 \pm 3.6$)/s \citep{Helms2009}) as is the the apparent exchange rate (Eq.~\eqref{eq:R1cross}): $R_\text{x}^{a} = (16.1 \pm 1.2)$/s compared to ($23 \pm 4$)/s \citep{Stanisz.2005}.
Notwithstanding, our analysis matches previous findings that the constraint $T_1^s = T_1^f$ biases $R_\text{x}$ to larger values \citep{Helms2009}.

Supporting Figs.~\ref{fig:InVivo_cf_Models}, \ref{fig:InVivo_MS}, and \ref{fig:ROI_cf_Models} compare the unconstrained qMT estimates to constrained fits of the same hybrid-state data. Most constrained estimates match neither the unconstrained estimates nor literature values.

\paragraph{Gray matter}
An ROI analysis of the cortical gray matter, averaged over all healthy volunteers, reveals trends similar to the WM analysis: $T_1^f = (2.46 \pm 0.56)$s and $T_1^s = (0.42 \pm 0.40)$s differ substantially from one another. The apparent $T_1^{f,a} = (1.62 \pm 0.23)\text{s}$ is in line with the mono-exponential estimate ($1.82 \pm 0.11$)s measured by \citet{Stanisz.2005}.

As expected, the estimated $m_0^s = 0.098 \pm 0.026$ is both smaller than that measured in WM and similar to the literature value $m_0^s = 0.086$ (derived from $\tilde{m}_0^s = 0.094$) estimated with the unconstrained MT model \citep{Helms.2004}.
The estimated $m_0^{s,a} = 0.071 \pm 0.051$ is in line with literature values based on a constrained MT model ($0.050 \pm 0.005$ \citep{Stanisz.2005}), though noise amplification resulting from Eq.~\eqref{eq:m0s_Taylor} limits the value of this comparison.

The estimated $R_\text{x} = (14.0 \pm 3.1)$/s as well as $R_\text{x}^a = (16.4 \pm 3.4)$/s of GM are, similarly to WM, lower than literature values that are based on a constrained MT model (($40 \pm 1$)/s \citep{Stanisz.2005}).

\begin{figure}[tb]
	\centering
	\ifOL
		\includegraphics[]{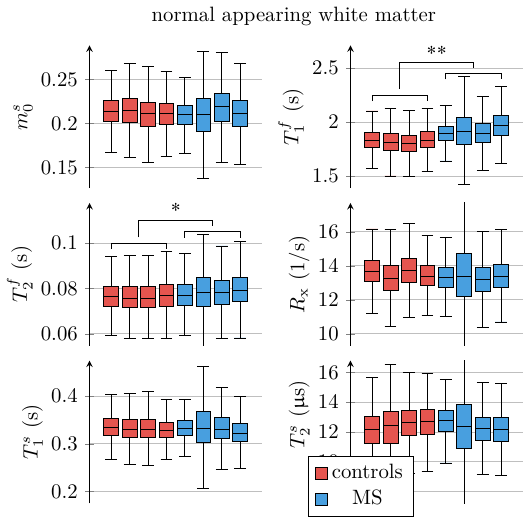}
	\else
		\input{Figures/ROI_WM_allParam.tex}
	\fi
	\caption{ROI analysis of the unconstrained qMT model's parameters, pooled over all normal-appearing white matter (NAWM) voxels in each of the 4 individuals with MS and the 4 controls.
		The markers $*$ and $**$ indicate statistically significant differences at the $p < 0.05$ and $p < 0.01$ levels. We note that the panels for $m_0^s$, $R_1^f$, and $R_\text{x}$ are a repetition of the ones in Fig.~\ref{fig:ROI_apparent}.}
	\label{fig:ROI_WM}
\end{figure}

\subsection{MS pathology}
\subsubsection{Normal-appearing white and gray matter}
Fig. \ref{fig:ROI_WM} compares all 6 unconstrained qMT parameters in an ROI spanning the entire NAWM between individuals with RRMS and healthy controls. We observe the most distinct differences in $T_1^f$: the median $T_1^f$ across the NAWM of each MS subject averaged over all participants with MS was 98ms larger than in controls ($p<0.01$).
By comparison, the apparent $T_1^{f,a}$ (Eq.~\eqref{eq:R1a}) differs only by 19ms ($p < 0.05$; cf. Fig.~\ref{fig:ROI_apparent}).

In NAWM, the median $T_2^f$ of each MS subject averaged over all participants with MS was 2.1ms larger than in controls ($p<0.05$).
When analyzing all unconstrained qMT parameters for the ROIs listed in Tab. \ref{tab:ROImean}, we found statistically significant changes of
\begin{itemize}
	\setlength{\parskip}{0pt}
	\setlength{\itemsep}{0pt plus 1pt}
	\item $T_1^f$, $T_2^f$, and $T_1^s$ in the anterior corpus callosum ($p<0.01$, $p<0.05$, $p<0.05$);
	\item $T_1^f$ in the posterior corpus callosum ($p<0.01$);
	\item $T_1^f$ in the cortical GM ($p < 0.05$);
	\item $T_1^f$ and $T_2^f$ in the caudate ($p<0.05$, $p<0.05$);
	\item $T_1^f$ in the pallidum ($p < 0.05$);
	\item $T_1^f$ in the putamen ($p < 0.05$).
\end{itemize}
With the constrained MT model, we only found significant differences in $T_1^{f,a}$ of the putamen ($p < 0.05$).
Fig.~\ref{fig:ROI_R1f} depicts $T_1^f$ for the ROIs above.

\begin{figure}[tb]
	\centering
	\ifOL
		\includegraphics[]{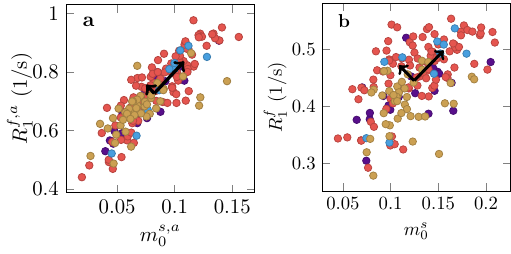}
	\else
		\begin{tikzpicture}[scale = 0.9]
    \begin{axis}[
        width=\columnwidth*0.4,
        height=\columnwidth*0.4,
        scale only axis,
        colormap={bw}{color(0cm)=(NYUpurple); color(1cm)=(Pastrami); color(2cm)=(TheLake); color(3cm)=(SpicyMustard)},
        xlabel={$m_0^{s,a}$},
        xticklabel style={/pgf/number format/fixed, /pgf/number format/precision=2},
        ylabel={$R_1^{f,a}$ (1/s)},
        ylabel style={yshift=-0.175cm},
        set layers, mark layer=axis tick labels,
        name=R1a,
        ]

        \addlegendimage{empty legend}
        \addlegendimage{empty legend}
        \addlegendimage{legend image code/.code={\draw[fill=TheLake,      draw=black, fill opacity = 1] (0cm,0cm) circle (0.07cm);}}
        \addlegendimage{legend image code/.code={\draw[fill=SpicyMustard, draw=black, fill opacity = 1] (0cm,0cm) circle (0.07cm);}}
        \addlegendimage{legend image code/.code={\draw[fill=Pastrami,     draw=black, fill opacity = 1] (0cm,0cm) circle (0.07cm);}}
        \addlegendimage{legend image code/.code={\draw[fill=NYUpurple,    draw=black, fill opacity = 1] (0cm,0cm) circle (0.07cm);}}

        \addplot +[only marks, scatter, scatter src=explicit] table [x=m0a, y=R1a_1-s, meta=subj]{Figures/Lesions.txt};

        \draw[black, ultra thick, ->] (0.08231230933751374, 0.7273975488117767) -- (0.10820524224119345, 0.8365798205823924);
        \draw[black, ultra thick, ->] (0.08231230933751374, 0.7273975488117767) -- (0.07523424840911473, 0.7572434830197016);

        \node[anchor=north west] at (rel axis cs: 0.05, .975)  {\textbf{a}};
    \end{axis}

    \begin{axis}[
            width=\columnwidth*0.4,
            height=\columnwidth*0.4,
            scale only axis,
            colormap={bw}{color(0cm)=(NYUpurple); color(1cm)=(Pastrami); color(2cm)=(TheLake); color(3cm)=(SpicyMustard)},
            xlabel={$m_0^s$},
            xticklabel style={/pgf/number format/fixed, /pgf/number format/precision=2},
            ylabel={$R_1^f$ (1/s)},
            ylabel style={yshift=-0.15cm},
            name=R1f,
            at=(R1a.right of south east),
            anchor=left of south west,
            set layers, mark layer=axis tick labels,
        ]

        \addplot +[only marks, scatter, scatter src=explicit] table [x=m0s, y=R1f_1-s, meta=subj]{Figures/Lesions.txt};

        \draw[black, ultra thick, ->] (0.12464072768177298, 0.4448468807765414) -- (0.15560601816267156, 0.49730546973475326);
        \draw[black, ultra thick, ->] (0.12464072768177298, 0.4448468807765414) -- (0.10879189205761103, 0.4716965418183793);

        \node[anchor=north west] at (rel axis cs: 0.05, .975)  {\textbf{b}};
    \end{axis}
\end{tikzpicture}
	\fi
	\caption{Analysis of longitudinal relaxation in lesions pooled across all 4 participants with MS where each color corresponds to one individual. \textbf{a} The median size of the apparent semi-solid spin pool $m_0^{s,a}$ vs the median apparent relaxation rate $R_1^{f,a}$. \textbf{b} Median $m_0^s$ vs $R_1^f$ as measured with the unconstrained MT model. The black arrows visualize the scaled eigenvectors of a PCA that quantify the independent variability in the respective model.
	}
	\label{fig:Lesions}
\end{figure}
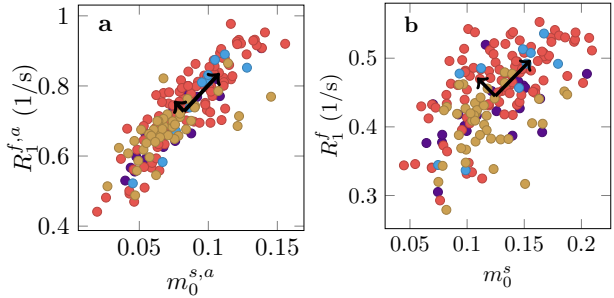

\subsubsection{MS lesions}
\begin{figure*}[tb]
	\centering
	\ifOL
		\includegraphics[]{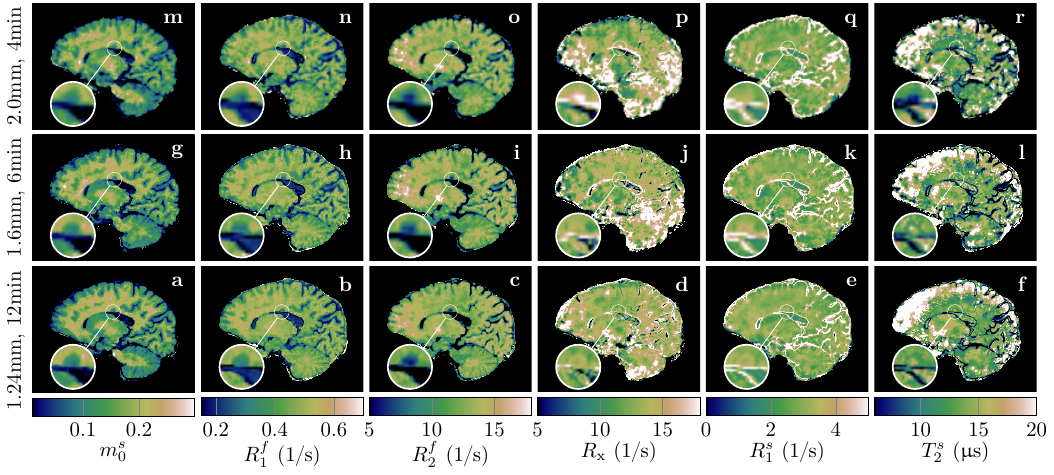}
	\else
		\input{Figures/In_Vivo_Resolution.tex}
	\fi
	\caption{
		Quantitative MT maps of an individual with MS. The rows compare different (isotropic) effective resolutions that require different scan times. All scans were acquired with full brain coverage. The magnifications show a lesion to highlight the resolution differences and the consequent partial volume effects.}
	\label{fig:InVivo_Resolution}
\end{figure*}

In MS lesions, we observe a substantial reduction of $m_0^s$ (Figs.~\ref{fig:InVivo_Resolution} and \ref{fig:InVivo_MS}) and $m_0^{s,a}$ (Fig.~\ref{fig:InVivo_MS_R1a}) relative to the NAWM, consistent with the expected demyelination.

When jointly analyzing $T_1^{f,a}$ and $m_0^{s,a}$ across all MS lesions using principal component analysis (Fig.~\ref{fig:Lesions}), we find that the first component explains 93\% of the variability.
By comparison, only 79\% of variability is explained by the first component for the unconstrained model, suggesting an increase in independent information across the qMT parameters. This might be beneficial in understanding the various biophysical processes contributing to disease, which we elaborate on in the Discussion.

\subsection{Rapid qMT imaging} \label{sec:ResultsRapidResolution}
All data described thus far were acquired with 1.24mm isotropic resolution and 36min scan time. To gauge the potential of our qMT approach for more clinically feasible scan times, we scanned an individual with MS with different resolutions and scan times. With 1.24mm isotropic resolution and 12min scan time, we observe overall good image quality despite slightly increased blurring and noise compared to the 36min scan (cf. the cerebellum in Fig. \ref{fig:InVivo_ctrl}f to the one in Fig. \ref{fig:InVivo_Resolution}a).
With 1.6mm isotropic nominal resolution and 6min scan time, we observe similar image quality besides the reduced resolution, and the same is true for 2.0mm isotropic in 4min.

\section{Discussion} \label{sec:Discussion}
\subsection{Unlocking unconstrained qMT imaging} \label{sec:DiscussionUnlocking}
Many pulse sequences are sensitive to $R_1^f$ and $R_1^s$. Using SIR \citep{Dortch.2018} as an example, this can be illustrated with normalized CRB values (cf. supporting Tab.~\ref{tab:CRB}) under the assumption that only the respective parameter is unknown:
norm. $\text{CRB}(R_1^f) \approx 25\text{s}$
and
norm. $\text{CRB}(R_1^s) \approx 20\text{s}$.
However, when considering $M_0$, $m_0^s$, $R_1^f$, $R_1^s$, and $R_x$ as unknown, the CRB increase to $192,160\text{s}$ and $117,183\text{s}$, respectively.
This difference can be understood with the geometric interpretation of the CRB that was introduced by \citet{Scharf.1993}:
The CRB of a single unknown parameter is simply the inverse squared $\ell_2$-norm of the signal's derivative wrt. respective parameter.
In the case of multiple unknowns, the CRB is given by the inverse squared $\ell_2$-norm of the signal's \textit{orthogonalized} derivative, i.e. after removing all components that are parallel to another gradient.
Consequently, the key to a low CRB and, ultimately, a stable fit is to disentangle the individual derivatives such that their orthogonalized components are large.

The SIR sequence maps a bi-exponential recovery after a $T_2$-selective inversion pulse. The relaxation curve is characterized by 5 parameters, which sets an upper limit for the number of model parameters.
An additional spin disturbance increases the number of observations, which opens the door for the estimation of additional parameters.
As illustrated above, however, the key to a \textit{stable} estimation of additional parameters is to disentangle the signal's derivatives wrt. the unknown parameters.

The proposed hybrid-state approach resembles SIR in the $T_2$-selective inversion pulse but adds a train of RF pulses for additional spin disturbances.
The large number of RF pulses provides many degrees of freedom to optimize the spin trajectory for maximum disentanglement of all derivatives, as captured by the CRB.

The choice between models with different numbers of parameters generally entails a variance-bias trade-off.
For the experimental design described in this paper, which was optimized for unconstrained qMT imaging, the CRB values associated with an unconstrained model, compared to a constraint model ($R_1^s$ fixed to 2s), are increased only by a factor of 2 in $R_1^f$, which corresponds to an SNR decrease of $\sqrt{2}$, assuming that the CRB is a tight bound.
For all other parameters, the CRB increases by a factor smaller than 1.03, which highlights the effectiveness of the disentanglement.
Comparing the CRB values to experimental designs that are optimized for constraint qMT (supporting Tab.~\ref{tab:CRB} vs. Tab.~1 in \citet{Assländer.2024}), the largest SNR penalty is a factor of 3 in $R_1^f$ (CRB increase by a factor of 9) in comparison to SIR with a 4-parameter model ($M_0$, $m_0^s$, $R_1^f$, and a $B_1^+$ proxy \citep{Cronin2020}) and much less in all other parameters.
These comparisons confirm the feasibility of unconstrained qMT with 9 parameters (plus a complex phase) with a manageable SNR penalty.

\subsection{Constrained vs. unconstrained MT models}
Our data confirms previous reports that estimates $T_1^s \ll T_1^f$ for white matter at 3T \citep{Helms2009,Gelderen2016,Manning2021,Samsonov2021,Zaiss.2022}.
We show that this finding has substantial implications for the estimation of the other model parameters.
With a Taylor expansion, we show that $T_1^f$ and $m_0^s$ are underestimated if $T_1^s = T_1^f$ is assumed (Section~\ref{sec:TheoryBiexpModel}) and a comparison of our experimental data to the literature confirms this finding.
Section~\ref{sec:TheoryBiexpModel} also highlights that the finding $T_1^s \ll T_1^f$ implies that MT drives the observed longitudinal relaxation, not just immediately following RF irradiation but throughout the MR experiment: in such a spin system continuous magnetization transfer to the semi-solid spin pool is a key driver of the apparent $T_1$ relaxation.
This stands in contrast to most MT literature, which assumes that the MT effect is mostly during RF irradiation and that, once the longitudinal magnetization of the two pools approaches each other (which happens at the time scale $T_\text{x}=1/R_\text{x}\approx50$ms), they relax independently.

Inserting unconstrained estimates of qMT parameters in white matter (Tab.~\ref{tab:ROImean}) into Eq.~\eqref{eq:Longitudinal} results in $T_1^{f,a} = 1/R_1^{f,a} \approx 0.94$s (supporting Tab.~\ref{tab:ROImean_apparent}), which is consistent with mono-exponential estimates reported in the literature ($T_1^{f,a} \approx 1.084$s \citep{Stanisz.2005}).
This concordance is expected for experiments with $z^f/m_0^f \approx z^s/m_0^s$, which can be achieved in inversion recovery experiments either by inverting both spin pools with a short RF-pulse ($T_\text{RF} \ll T_2^s$)---which is not feasible in vivo, but was done by \citet{Stanisz.2005} in their NMR experiments---or by choosing inversion times that fulfill $T_\text{I} \gg T_\text{x}$.
Our pulse sequence does not fulfill this condition, which explains the deviating $T_1^{f,a} \approx 1.429$s when fitting a mono-exponential model to our data (Supporting Fig.~\ref{fig:InVivo_cf_Models}).

\subsection{Myelin as a contrast agent} \label{sec:DiscussionMyelin}
\citet{Koenig1990} suggested that myelin is the primary source of GM-WM contrast in $T_1$-weighted MRI (Fig. \ref{fig:InVivo_ctrl}c,d), an observation that extends to $R_1^{f,a}$ maps (Fig.~\ref{fig:InVivo_ctrl_R1a}).
In an unconstrained MT model, the pronounced GM-WM contrast shifts from $R_1^f$ to $m_0^s$ (Fig. \ref{fig:InVivo_ctrl}e,f).
This observation refines the finding of \citet{Koenig1990} by identifying MT as the primary mechanism that generates the observed GM-WM contrast.
However, we also observe a subtle GM-WM contrast in $T_1^f$, which may suggest that myelin also facilitates direct longitudinal relaxation of the free spin pool beyond MT, possibly by interactions between water protons and the local magnetic field of myelin (or macromolecules in general, see \citet{Gossuin.2000}).
This observation is consistent with our phantom experiments (Fig.~\ref{fig:Phantom}), where $R_1^f$ was found to be linearly dependent on the BSA concentration.

\subsection{Iron as a contrast agent}
$R_1^f$ of the pallidum was shorter than that of all other ROIs analyzed in this study (Tab.~\ref{tab:ROImean}).
Since iron is known to accumulate in the pallidum in the form of ferritin, this suggests a sensitivity of $R_1^f$ to iron, which matches the reports by \citet{Vymazal.1999} and \citet{Samsonov2021}.
Supporting Fig.~\ref{fig:Iron} fits $R_1^f$ as a function of the iron concentration in each ROI as taken from the literature, revealing a linear dependency ($R^2 = 0.94$).
Repeating the same analysis for the transversal relaxation rate $R_2^f$ reveals a much clearer linear dependency ($R^2 = 0.9998$), suggesting that $T_2^f$ is more sensitive and specific to iron than $R_1^f$, in line with previous reports by \citet{Schenker.1993,Vymazal.1999,Gossuin.2000}.

\subsection{Myelin water as a confounding factor}
Our WM estimates of $T_2^f$ deviate from previous reports \citep{Stanisz.2005}. A possible explanation is that our model neglects contributions from myelin water (MW)---or water trapped between the myelin sheaths---that has a characteristic $T_2^{\text{MW}} \approx 10$ms \citep{Mackay1994a}. MW exchanges magnetization with myelin's macromolecular pool as well as the larger intra-/extra-axonal water pool, where the former exchange is faster than the latter \citep{Stanisz1999,Manning2021}.
A saturation of the semi-solid pool could, thus, result in a saturation of the MW pool and, ultimately, its suppression.
A subsequent estimate of the observed $T_2^f$---which comprises both the intra-/extra-axonal water pool and MW pool---would thus be dominated by the former and result in higher observed $T_2^f$ values. By contrast, a CPMG sequence starts from thermal equilibrium and has more pronounced contributions from the MW pool, resulting in shorter observed $T_2^f$. However, a more detailed analysis is needed for a thorough understanding of these observed deviations.

\subsection{Unconstrained qMT in multiple sclerosis}
Supporting Fig. \ref{fig:InVivo_MS}u highlights four MS lesions with a hypointense appearance in the MP-RAGE.
Our data suggests that this hypointensity is primarily driven by a reduction of $m_0^s$, which was observed in most examined lesions (Fig.~\ref{fig:Lesions}b).
By contrast, we find that changes in the NAWM are primarily driven by $T_1^f$ (Fig.~\ref{fig:ROI_apparent}). This juxtaposition of the different sources of contrast changes highlights the complexity of longitudinal relaxation in biological tissue.

In histology, MS lesions exhibit substantial heterogeneity in terms of varying degrees of remyelination, axonal damage, inflammation, and gliosis \citep{Lassmann.2018}.
Fig.~\ref{fig:Lesions} suggests that unconstrained qMT can delineate more independent information as compared to constrained qMT.
Future work will aim to identify links between qMT parameters and pathological variability in MS lesions.

Another goal of this paper was to gauge the sensitivity of unconstrained qMT to subtle changes in normal-appearing WM and GM that are not easily detectable with established (contrast-based) clinical sequences.
We observed statistically significant deviations of $T_1^f$ between individuals with MS and healthy controls, in particular, in the NAWM, which aligns with previous studies that performed mono-exponential $T_1$-mapping \citep{Vrenken.2006,Vrenken.20066th,Vrenken.2006sph}.
Moreover, we found statistically significant deviations of $T_1^f$ in subcortical GM structures.
An analysis of NAWM in individuals with MS always bears the risk of contaminating the results with an incomplete exclusion of MS lesions or by voxels close to lesions. However, we have two reasons to believe that lesions and their surrounding tissue do not drive the observed changes in $R_1^f$. First, we predominantly observe changes in $m_0^s$ in lesions, while $m_0^s$ changes in NAWM are much less pronounced. Second, \citet{Vrenken.2006sph} demonstrated that the magnetization transfer ratio in NAWM changes with the distance to an MS lesion, but their mono-exponential $T_1$ estimates do not.
Another limitation of this study is the small number of participants, which does not allow for adjustments, e.g., of the age difference between the two cohorts.
Therefore, larger studies are needed to confirm this result.

\subsection{Rapid, high-resolution qMT imaging} \label{sec:DiscussionResolution}
A major goal of this paper is to demonstrate the feasibility of unconstrained qMT imaging on a voxel-by-voxel basis.
With a hybrid-state pulse sequence, we were able to extract unconstrained qMT maps with 1.24mm, 1.6mm, and 2.0mm isotropic resolution from 12min, 6min, and 4min scans, respectively.
To the best of our knowledge, the presented maps are the first voxel-wise fits using an unconstrained MT model.
However, we do observe a subtle blurring in our qMT maps compared to the MP-RAGE.
The most likely cause is the smaller k-space coverage of the koosh-ball trajectory in comparison to a Cartesian trajectory: the koosh-ball trajectory with a nominal resolution of 1.0mm samples only the inner sphere of the 1.0mm k-space cube, similar to \textit{elliptical scanning}, while the MP-RAGE samples the entire cube.
We account for the reduced k-space coverage using the ``effective'' resolution of 1.24mm.
Undersampling, regularized reconstruction, and incomplete motion correction might cause additional blurring.
On the flip side, our image reconstruction models the spin dynamics, alleviating relaxation-induced blurring that is more prevalent in approaches like MP-RAGE \citep{Mugler1990} or RARE \citep{Hennig1986}.

\subsection{Future Directions} \label{sec:DiscussionOutlook}
Our ongoing work includes clinical validation as well as efforts for further scan time reductions and improvements in resolution.
To this end, we aim to replace the current RF pattern, which is a concatenation of separate optimizations, with a joint optimization of all unconstrained qMT parameters. Further, we are exploring more efficient k-space trajectories.
Last, we anticipate that studies with the current pulse sequence will help identify the most clinically meaningful parameters. This information can then be fed back to our numerical optimization framework to optimize pulse sequences for more efficient estimation of these parameters.
Optimizations of the sequence for particular parameters can be achieved using the employed CRB-based framework without imposing constraints on the parameters.

\section{Conclusion}
Our study builds on the work of \citet{Helms2009,Gelderen2016,Manning2021,Samsonov2021,Zaiss.2022}, who pioneered unconstrained fitting with Henkelman's two-pool magnetization transfer model.
By utilizing the encoding power of the hybrid state \citep{Assländer.2019}, we improved the sensitivity of the MRI data to the model's parameters, enabling an unconstrained fit of the MT model to each voxel separately.
Our results confirm previous observations of the substantially different longitudinal relaxation times of the free and semi-solid spin pools.
The results also suggest a potential clinical value of unconstrained qMT for individuals with MS via the detection of changes in the NAWM and the characterization of MS lesions.

\appendix
\renewcommand{\theequation}{A.\arabic{equation}}
\renewcommand{\thefigure}{A.\arabic{figure}}

\section{Eigendecomposition of longitudinal relaxation} \label{sec:AppendixEigendecomposition}
The eigenvector associated with thermal equilibrium is
\begin{equation}
	\mathbf{v}_e = \begin{pmatrix} m_0^f \\ m_0^s \\ 1 \end{pmatrix}.
\end{equation}
The eigenvector associated with the apparent relaxation rate is
in the approximation of a Taylor expansion up to the linear term
\begin{equation}
	\mathbf{v}_1 \approx \begin{pmatrix} m_0^f \left(1 + \frac{R_1^s-R_1^f}{R_\text{x}}\right) \\ m_0^s \\ 0 \end{pmatrix}.
\end{equation}
In the same approximation, the eigenvector associated with the cross-relaxation term is given by
\begin{equation}
	\mathbf{v}_\text{x} \approx \begin{pmatrix} R_1^s-R_1^f-R_\text{x} \\ R_\text{x} \\ 0 \end{pmatrix}.
\end{equation}
We note that all three eigenvalues are uniquely defined up to a scaling factor.

In an inversion recovery experiment, the dynamics of the $z$-magnetization of the two-pool system are given by
\begin{equation}
	\mathbf{m}(t) = c_e \mathbf{v}_e - c_\text{x} \mathbf{v}_\text{x} \exp(- R_\text{x}^a t) - c_1 \mathbf{v}_1 \exp(- R_1^{f,a} t).
	\label{eq:mtime}
\end{equation}
Assuming the thermal equilibrium $z^f(+\infty) = m_0^f$ and $z^s(+\infty) = m_0^s$ and a perfect selective inversion-recovery (SIR) experiment \citep{Gochberg.2003} with the initial conditions $z^f(0) = -m_0^f$ and $z^s(0) = m_0^s$, i.e., an inversion of the free pool with no effect on the semi-solid pool, we can calculate coefficients $c_{e,1,\text{x}}$. Defining $c_{\text{x}}^f := c_{\text{x}} \mathbf{v}_\text{x}^{(1)}$, where $\mathbf{v}_\text{x}^{(1)}$ denotes the first element of $\mathbf{v}_\text{x}$, we approximate this coefficient $c_\text{x}^f$ by the Taylor expansion
\begin{equation}
	c_\text{x}^f \approx 2 m_0^f m_0^s \left(1 - \frac{2 m_0^f (R_1^s - R_1^f)}{R_\text{x}}\right).
	\label{eq:cxf_def}
\end{equation}
If we were to assume $(R_1^s - R_1^f) \ll R_\text{x}$, we would measure the \textit{apparent} pool sizes:
\begin{equation}
	c_\text{x}^f := 2 m_0^{f,a} m_0^{s,a}.
	\label{eq:cxf_apparent}
\end{equation}
We note that \citet{Gochberg.2003} derived $c_\text{x}^f \approx 2 m_0^{s,a}$ based on slightly different approximations.
Combining Eqs.~\eqref{eq:cxf_def} and \eqref{eq:cxf_apparent} and approximating $m_0^{f,a} \approx m_0^f$ results in Eq.~\eqref{eq:m0s_Taylor}.

We note that the parameter $c_\text{x}^f$ can be estimated experimentally by fitting the following bi-exponential model to an inversion-recovery experiment \citep{Gochberg.2003}:
\begin{equation}
	z^f(t) = m_0^f - c_\text{x}^f \exp(- R_\text{x}^a t) - c_1^f \exp(- R_1^{f,a} t),
	\label{eq:zf}
\end{equation}
which corresponds to the first row of Eq.~\eqref{eq:mtime}.

\section{Data availability statement} \label{sec:data_availability}
In order to promote reproducibility, we provide the latest version (v0.8.0, DOI:10.5281/zenodo.7433494) of the sequence optimization and signal simulation source code on \url{https://github.com/JakobAsslaender/MRIgeneralizedBloch.jl}. They are written in the open-source language \href{https://julialang.org}{Julia} and registered to the package manager as ``MRIgeneralizedBloch.jl." The package documentation and tutorials can be found at \url{https://JakobAsslaender.github.io/MRIgeneralizedBloch.jl}. The tutorials render the code in HTML format with interactive figures and links to Jupyter notebooks that can be launched in a browser without local installations using \textit{binder}.

We also provide the source code of the image reconstruction package at \url{https://github.com/JakobAsslaender/MRFingerprintingRecon.jl}. For the presented data, we used v0.6.0.

The qMT maps of all participants are available at \url{https://doi.org/10.5281/zenodo.7492581}.

\section{Author contributions}
\textbf{Jakob Assl\"ander: }Conceptualization; Data curation; Formal analysis; Funding acquisition; Investigation; Methodology; Project administration; Software; Supervision; Visualization; Writing - original draft; Writing - review \& editing.
\textbf{Andrew Mao: }Funding acquisition; Software; Writing - review \& editing.
\textbf{Elisa Marchetto: } Motion Correction; Writing - review \& editing.
\textbf{Erin S Beck: }Methodology (lesion segmentation); Conceptualization; Writing - review \& editing.
\textbf{Francesco La Rosa: }Methodology (lesion segmentation); Writing - review \& editing.
\textbf{Robert W Charlson: }Resources (patient recruitment); Writing - review \& editing.
\textbf{Timothy Shepherd: }Conceptualization; Writing - review \& editing.
\textbf{Sebastian Flassbeck: }Data curation; Investigation; Software; Writing - review \& editing.

\section{Funding}
This research was supported by the
NIH/NINDS grant R01~NS131948, NIH/NIBIB grant R21~EB027241, and was performed under the rubric of the Center for Advanced Imaging Innovation and Research, an NIBIB Biomedical Technology Resource Center (NIH~P41 EB017183).
AM acknowledges support from the NIH/NIA (T32~GM136573 and F30~AG077794). FLR is supported by the Swiss National Science Foundation (SNF) Postdoc Mobility Fellowship (P500PB\_206833).

\section{Declaration of Competing Interests}
None of the authors have any competing interests to declare.

\bibliographystyle{elsarticle-harv}
\bibliography{library}%

\makeatletter \@input{Supporting_Information.aux.tex} \makeatother

\end{document}


\begin{frontmatter}
	\title{\Papertitle}

	\author[1,2]{Jakob Assl\"ander\footnotemark\textsuperscript{,}}
	\author[1,2,3]{Andrew Mao}
	\author[1,2]{Elisa Marchetto}
	\author[4]{Erin S Beck}
	\author[4]{Francesco La Rosa}
	\author[5]{Robert W Charlson}
	\author[1]{Timothy M Shepherd}
	\author[1,2]{Sebastian Flassbeck}

	\affiliation[1]{
		organization={Center for Biomedical Imaging, Dept. of Radiology, New York University School of Medicine},
		addressline={650 1st Avenue},
		city={New York},
		postcode={10016},
		state={NY},
		country={USA}
	}

	\affiliation[2]{
		organization={Center for Advanced Imaging Innovation and Research (CAI2R), Dept. of Radiology, New York University School of Medicine},
		addressline={650 1st Avenue},
		city={New York},
		postcode={10016},
		state={NY},
		country={USA}}

	\affiliation[3]{
		organization={Vilcek Institute of Graduate Biomedical Sciences, New York University School of Medicine},
		addressline={550 1st Avenue},
		city={New York},
		postcode={10016},
		state={NY},
		country={USA}}

	\affiliation[4]{
		organization={Corinne Goldsmith Dickinson Center for Multiple Sclerosis, Department of Neurology, Icahn School of Medicine at Mount Sinai},
		addressline={5 East 98th Street},
		city={New York},
		postcode={10029},
		state={NY},
		country={USA}}

	\affiliation[5]{
		organization={Department of Neurology, New York University School of Medicine},
		addressline={240 E 38th Street},
		city={New York},
		postcode={10016},
		state={NY},
		country={USA}}


\end{frontmatter}

\footnotetext[1]{corresponding author: Jakob Assl\"ander, Center for Biomedical Imaging, Department of Radiology, New York University School of Medicine, 650 1st Avenue, New York, NY 10016, USA.\\ jakob.asslaender@nyumc.org}

\footnotetext[2]{abbreviations:
	WM: white matter,
	GM: gray matter,
	MS: multiple sclerosis,
	(q)MT: (quantitative) magnetization transfer,
	CRB: Cram\'er-Rao bound,
	BSA: bovine serum albumin,
	NN: neural network,
	ROI: region of interest,
	NAWM: normal-appearing white matter,
	EDSS: expanded disability status scale,
	CC: corpus callosum,
	MP-RAGE: magnetization-prepared rapid gradient-echo,
	FLAIR: fluid-attenuated inversion recovery,
	MW: myelin water
}

\begin{figure*}[htb]
	\centering
	\ifOL
		\includegraphics[]{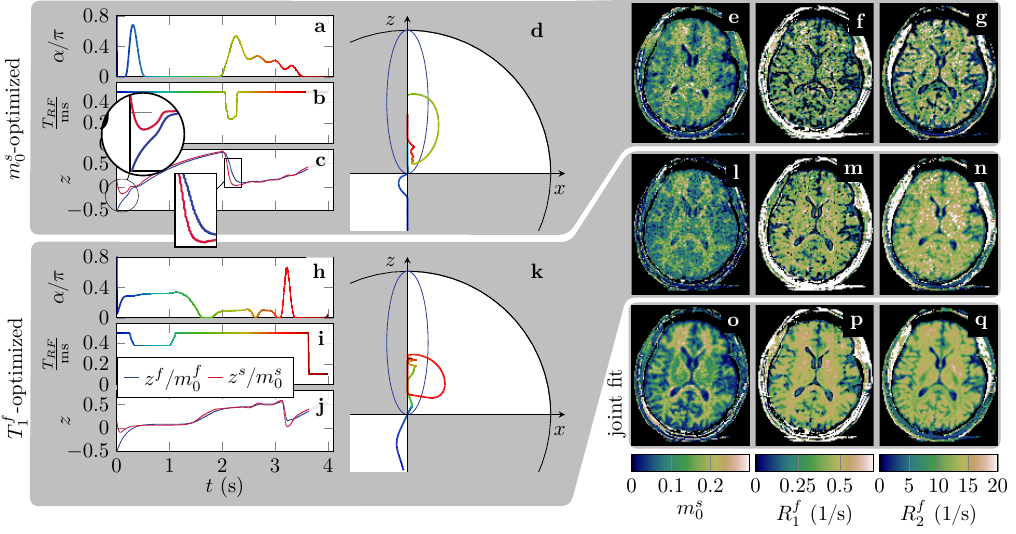}
	\else
		\input{Figures/SpinDynamics.tex}
	\fi
	\caption{Optimized RF-pulse trains, evoked spin dynamics, and corresponding in vivo parameter maps.
		\textbf{a,h} The flip angle $\alpha$ and
		\textbf{b,i} the pulse duration $T_\text{RF}$ control the spin dynamics.
		\textbf{c,j} The normalized $z$-magnetization of the two pools. The spherical and rectangular magnifications in (c) highlight segments that utilize a bi-exponential inversion-recovery \citep{Gochberg.2003} and saturation of the semi-solid spin pool \citep{Gloor2008}, respectively, which encode the semi-solid pool size $m_0^s$.
		\textbf{d,k} The dynamics of the free pool on the Bloch sphere with the steady-state ellipse in blue.
		\textbf{e-g} $m_0^s$ and the relaxation times of the free pool $T_1^f$ and $T_2^f$ that were estimated from a 6min scan with an $m_0^s$-optimized RF-pattern in comparison to \textbf{l-n} a pattern that was optimized for $T_1^f$ and \textbf{o-q} a joint fit of 6 measurements with RF-patterns that were optimized for different qMT parameters (36min scan time).
		We note that parameter maps are shown here to illustrate the trend, but we cannot expect a one-to-one match with the CRB values shown in Tab.~\ref{tab:CRB} as the CRB does not directly predict the noise variance for most practical estimators, but we rather use it as a proxy for the conditioning of the fitting problem (cf. Section~\ref{sec:MethodsNumericalOptimzation} in the main paper).
	}
	\label{fig:Spin_Dynamics}
\end{figure*}

\begin{table*}[htb]
	\centering
	\newcommand{\g}{\cellcolor[gray]{0.8}}
	\begin{tabular}{l|c|c|c|c|c|c|c|c||c}
		optimized for                                                         & $m_0^s$ & $R_1^f$ & $R_2^f$ & $R_\text{x}$ & $R_1^s$ & $T_2^s$ & joint    & concat. & concat. \\
		\hline
		$\text{CRB}(m_0^s     ) \cdot M_0^2/(m_0^s     \sigma)^2 \cdot T$ (s) & \g$47$  & $3837$  & $23397$ & $1576$       & $19593$ & $2169$  & \g$119$  & $99$    & $98$    \\
		$\text{CRB}(R_1^f     ) \cdot M_0^2/(R_1^f     \sigma)^2 \cdot T$ (s) & $2649$  & \g$91$  & $6564$  & $2608$       & $1147$  & $1674$  & \g$427$  & $237$   & $119$   \\
		$\text{CRB}(R_2^f     ) \cdot M_0^2/(R_2^f     \sigma)^2 \cdot T$ (s) & $969$   & $493$   & \g$15$  & $473$        & $652$   & $70$    & \g$172$  & $45$    & $44$    \\
		$\text{CRB}(R_\text{x}) \cdot M_0^2/(R_\text{x}\sigma)^2 \cdot T$ (s) & $21466$ & $17598$ & $52736$ & \g$185$      & $50744$ & $15552$ & \g$705$  & $449$   & $437$   \\
		$\text{CRB}(R_1^s     ) \cdot M_0^2/(R_1^s     \sigma)^2 \cdot T$ (s) & $24410$ & $13581$ & $20283$ & $58802$      & \g$264$ & $6928$  & \g$1263$ & $425$   & fixed   \\
		$\text{CRB}(T_2^s     ) \cdot M_0^2/(T_2^s     \sigma)^2 \cdot T$ (s) & $22270$ & $3708$  & $25178$ & $10160$      & $12876$ & \g$203$ & \g$717$  & $646$   & $628$   \\
	\end{tabular}
	\caption{
		Comparison of the Cram\'er-Rao bound (CRB) values (lower is better) between specialized optimizations for a single parameter, a concatenation of these 6 optimized RF patterns, and a joint optimization of all parameters.
		The objective of each optimization is highlighted in gray.
		The CRB values are normalized by the squared value of the parameter, the squared magnetization $M_0$, and the noise variance of the time series in a voxel $\sigma^2$, i.e., they reflect the inverse squared signal-to-noise ratio for a unit signal-noise variance.
		Further, they are normalized by the (simulated) scan time $T$, allowing for a fair comparison of the concatenated and the other patterns.
		Each optimization treats all biophysical parameters, as well as $\omega_z$, $B_1^+$, and the scaling factor $M_0$ as unknowns, i.e., we assume a 9-parameter fit, with the exception of the last row that treats $R_1^s$ as known for comparison.
	}
	\label{tab:CRB}
\end{table*}

\section{Numerical optimizations of the pulse train} \label{supsec:NumericalOptimizations}
The numerical optimizations of the pulse train resulted in smooth flip angle and $T_\text{RF}$ patterns. Fig. \ref{fig:Spin_Dynamics} sketches the RF patterns that were optimized for $m_0^s$ and $R_1^f$, respectively, along with the evoked spin trajectories.
We observe distinct patterns for the different optimization objectives (Fig.~\ref{fig:Spin_Dynamics}) that correspond to distinct CRB values (Tab.~\ref{tab:CRB}). For example, optimizing for $m_0^s$ resulted in a normalized CRB of 47s for $m_0^s$ and of 3837s for $R_1^f$, while the optimization for $R_1^f$ resulted in a normalized CRB of 2649s for $m_0^s$ and of 91s for $R_1^f$.
This difference in CRB values is in line with the noise levels in scans with each of the two RF patterns: the optimization for $m_0^s$ results in a comparably low noise level in $m_0^s$ (Fig. \ref{fig:Spin_Dynamics}e) and a comparably high noise level in $R_1^f$ (f),
while the optimization for $R_1^f$ results in the opposite noise characteristics (l vs. m).

As the optimizations aim to disentangle the effect of 9 overall parameters, the resulting spin trajectories are difficult to interpret.
Nonetheless, we can discern some features: for example, the $m_0^s$-optimized pattern starts with near-zero flip angles after the inversion pulse, which provokes a bi-exponential inversion recovery of the longitudinal magnetization (circular magnification in Fig.~\ref{fig:Spin_Dynamics}c) that encodes $m_0^s$ similar to the \textit{SIR} method proposed by \citet{Gochberg.2003}. The rectangular magnification in Fig.~\ref{fig:Spin_Dynamics} highlights a section of the spin dynamics in which large flip angles, combined with short $T_\text{RF}$, saturate the semi-solid spin pool, which resembles the encoding mechanism proposed by \citet{Gloor2008}.
This saturation maximizes the difference between pools, and this encoding mechanism is not as pronounced in the spin trajectory evoked by the $R_1^f$-optimized pattern, where the semi-solid spin pool plays a subordinate role.

Concatenating the 6 individual optimizations for $m_0^s$, $R_1^f$, $R_2^f$, $R_\text{x}$, $R_1^s$, and $T_2^s$, respectively, increases the CRB values (normalized by the scan time) compared to the optimized CRB value of each specialized RF-pattern, but provides overall low CRB values for all six parameters that also results in high-quality parameter maps (Fig. \ref{fig:Spin_Dynamics}o-q).
Interestingly, these CRB values are consistently lower than a joint optimization for all parameters. We note that the latter has to encode all parameters in a single 4s-long cycle. At the same time, the former utilizes 6 distinct RF patterns and has, thus, more degrees of freedom to induce different spin dynamics.
Given this result, we performed all experiments with concatenated scans that utilize the 6 individual optimizations.

\section{Noise analysis} \label{supSec:Noise}
\begin{figure}[htbp]
	\centering
	\ifOL
		\includegraphics[]{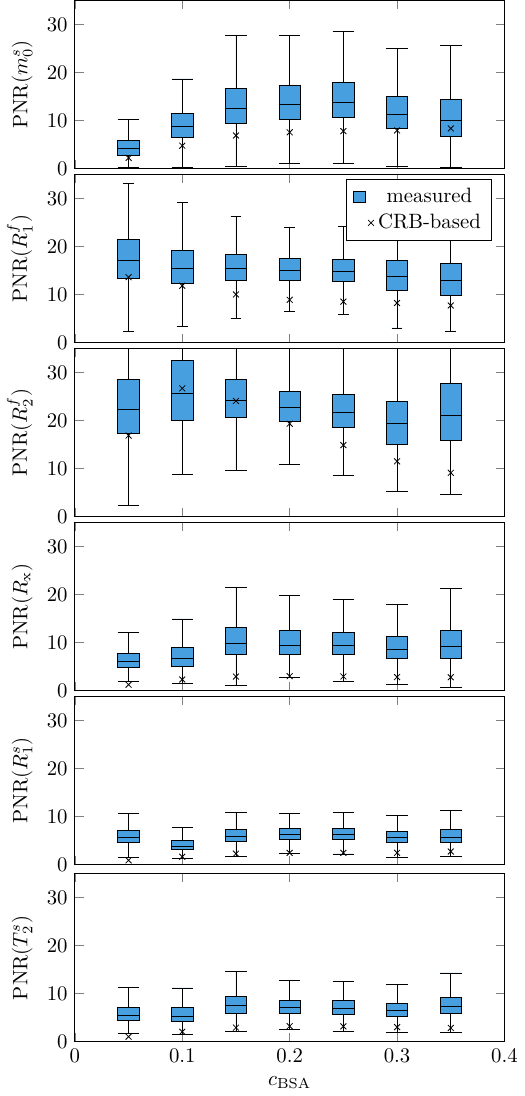}
	\else
		\input{Figures/Phantom_repeatability_CG.tex}
	\fi
	\caption{Parameter to noise ratio (PNR) of phantom scans at 2.9T. The PNR was measured voxel-by-voxel based on 25 repetitions of the same scan (reconstructed without any regularization) and the box plots show the inter-voxel variability of the estimated PNR.
		The x-markers identify the PNR as estimated by the median parameter estimates divided by the corresponding Cram\`er-Rao bound (CRB).}
	\label{fig:Phantom_PNRCG}
\end{figure}

In order to relate the Cram\`er-Rao bound to noise levels in the parameter maps, we measured a BSA phantom (cf. Section~\ref{sec:MethodsPhatom} in the main paper) 25 times for 12min with 1.24mm effective resolution. We reconstructed each dataset with the pipeline described in Section~\ref{sec:MethodsImageRecon}, but without the locally-low-rank regularization to minimize distortions of the noise statistics.
We calculated the parameter-to-noise ratio (PNR) voxel-by-voxel by dividing each parameter's estimate by the standard deviation over the 25 repetitions.
Fig.~\ref{fig:Phantom_PNRCG} visualizes the PNR-variability between different voxels of the same sample as box plots.

For comparison, we estimated the anticipated PNR based on the Cram\`er-Rao bound. For each BSA concentration, we used the median parameter estimates and the median noise variance in the coefficient images to calculate the CRB. The CRB-based PNR is then given by the median parameter estimate divided by the square root of the CRB (x-markers in Fig.~\ref{fig:Phantom_PNRCG}).
Comparing the experimental to the CRB-based PNR, we observe overall good correspondence. For virtually all parameters and samples, the CRB-based PNR lies within 1.5x the interquartile range, indicated by the whiskers.
The residual discrepancy can be explained by the CRB's role as a mere proxy for the PNR: the CRB is achieved only for an unbiased and fully efficient estimator.
Since neither non-linear-least square fitting nor the here-used neural-network fitting is strictly unbiased \citep{Newey.1994,Wu.1981}, the CRB can only serve as a proxy and Fig.~\ref{fig:Phantom_PNRCG} demonstrates sufficient concordance to render the CRB a useful metric for parameter estimation (cf. Section~\ref{sec:MethodsNumericalOptimzation} in the main paper).

\begin{figure}[htbp]
	\centering
	\ifOL
		\includegraphics[]{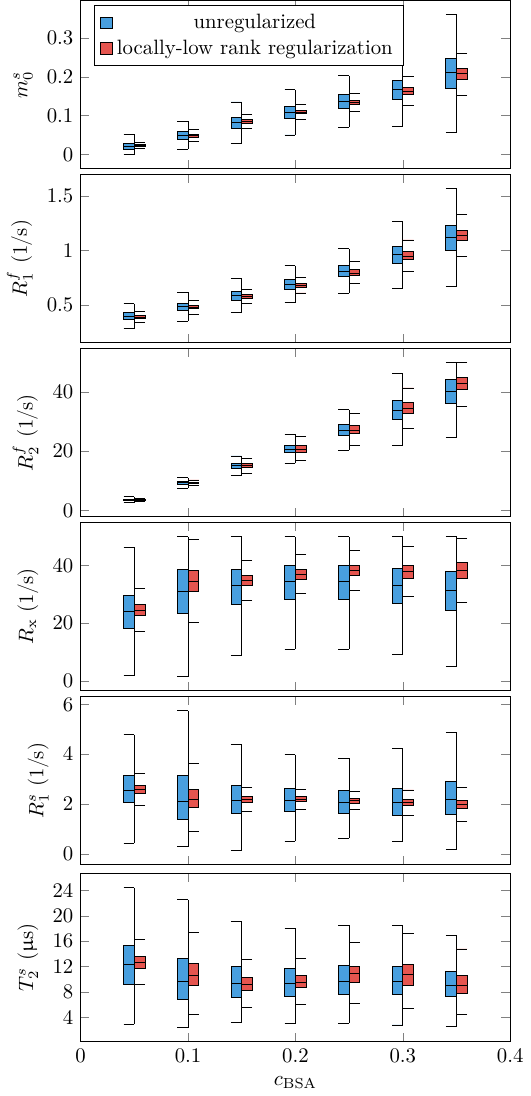}
	\else
		\input{Figures/Phantom_repeatability_LLRvCG.tex}
	\fi
	\caption{Comparison of parameter estimates between unregularized and regularized reconstructions at 2.9T. Each box plot analyzes the parameter estimates in a sample, pooled over all voxels in respective ROI and over 25 repetitions of the scan. The utilized regularization, including the regularization parameter, was identical to the reconstruction of the in vivo data shown in Fig.~\ref{fig:InVivo_Resolution}.
	}
	\label{fig:Phantom_repLLRvCG}
\end{figure}

To understand the effect of the employed locally-low-rank regularization on the parameter estimates, we repeated the phantom reconstructions with our default pipeline.
Fig.~\ref{fig:Phantom_repLLRvCG} compares the median estimates and their variability throughout each sample and the 25 repetitions (pooled together). The median values overall align well between the two reconstructions and we observe reduced variability in the estimates when utilizing regularization.

\begin{table}[htbp]
	\centering
	\begin{tabular}{l|c|c|c}
		          & $m_0^{s,a}$       & $T_1^{f,a}$ (s)   & $R_\text{x}^a$ (1/s) \\ \hline
		entire WM & $0.151 \pm 0.022$ & $0.941 \pm 0.069$ & $16.1 \pm 1.2$       \\
		ant. CC   & $0.174 \pm 0.029$ & $0.899 \pm 0.065$ & $15.8 \pm 1.9$       \\
		post. CC  & $0.173 \pm 0.033$ & $0.914 \pm 0.074$ & $15.9 \pm 2.1$       \\
		cort. GM  & $0.071 \pm 0.051$ & $1.62 \pm 0.23$   & $16.4 \pm 3.4$       \\
		Caudate   & $0.086 \pm 0.019$ & $1.389 \pm 0.080$ & $16.0 \pm 2.5$       \\
		Putamen   & $0.089 \pm 0.015$ & $1.271 \pm 0.096$ & $17.3 \pm 2.0$       \\
		Pallidum  & $0.125 \pm 0.022$ & $1.032 \pm 0.049$ & $18.3 \pm 1.9$       \\
		Thalamus  & $0.119 \pm 0.024$ & $1.23 \pm 0.13$   & $16.4 \pm 2.1$       \\
		Hippoc.   & $0.070 \pm 0.023$ & $1.62 \pm 0.16$   & $18.0 \pm 3.1$       \\
	\end{tabular}
	\caption{
		Region of interest (ROI) analysis of apparent qMT parameters (Eqs.~\eqref{eq:R1a}--\eqref{eq:m0s_Taylor}) in healthy controls. The ROIs were determined by segmenting the MP-RAGE images with the \textit{FreeSurfer} software after co-registering it to the qMT coefficient images. The values represent the mean and standard deviation of all voxels from 4 healthy participants.
	}
	\label{tab:ROImean_apparent}
\end{table}

\begin{figure*}[htbp]
	\centering
	\ifOL
		\includegraphics[]{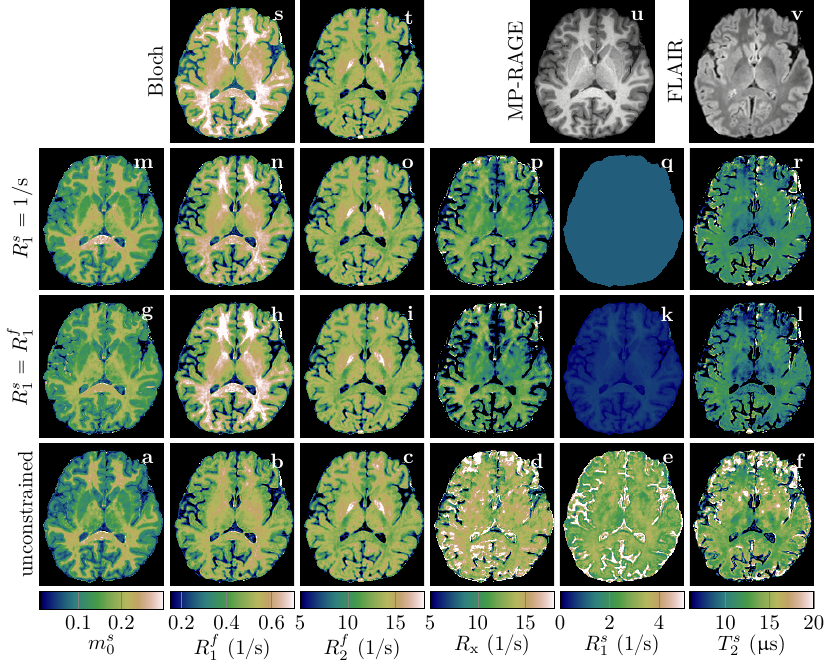}
	\else
		\input{Figures/In_Vivo_cfModels.tex}
	\fi
	\caption{Quantitative MT maps of a healthy participant fitted with the proposed unconstrained MT model (\textbf{a-f}), MT model constrained by $R_1^s = R_1^s$ (\textbf{g-l}) and $R_1^s = 1$/s (\textbf{m-r}), and the Bloch model (\textbf{s,t}).
		The fits with each model are based on separately trained neural networks.
		Note that \textbf{k} is a replica of \textbf{h} on a different color scale. MP-RAGE (\textbf{u}) and FLAIR (\textbf{v}) scans are provided for reference.}
	\label{fig:InVivo_cf_Models}
\end{figure*}

\begin{figure*}[htbp]
	\centering
	\ifOL
		\includegraphics[]{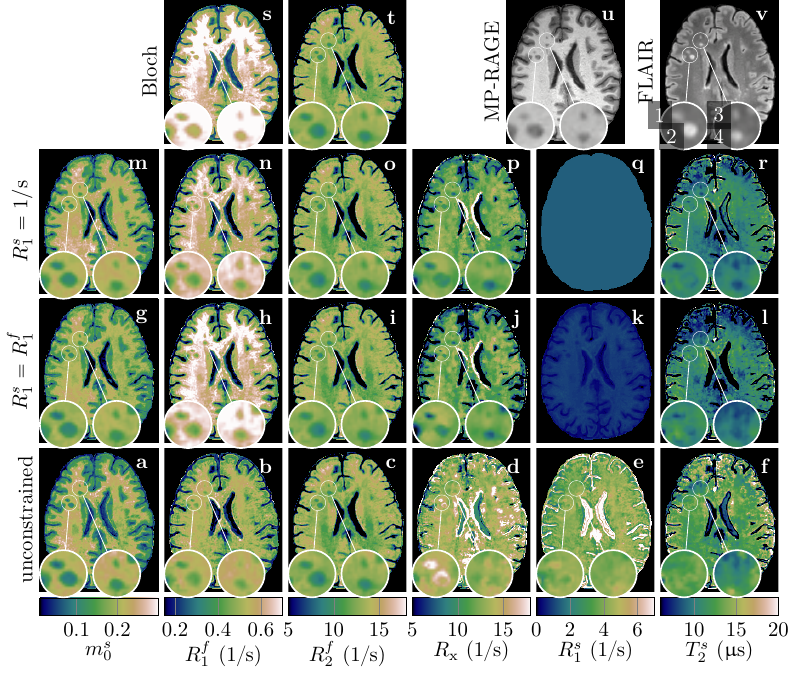}
	\else
		\input{Figures/In_Vivo_MS.tex}
	\fi
	\caption{Quantitative MT maps of an individual with MS fitted with the proposed unconstrained MT model (\textbf{a-f}), MT model constrained by $R_1^s = R_1^s$ (\textbf{g-l}) and $R_1^s = 1$/s (\textbf{m-r}), and the Bloch model (\textbf{s,t}).
		The fits with each model are based on separately trained neural networks.
		Note that \textbf{k} is a replica of \textbf{h} on a different color scale. MP-RAGE (\textbf{u}) and FLAIR (\textbf{v}) scans are provided for reference. The magnifications highlight four lesions (labeled in \textbf{v}) that have similar appearances on the FLAIR and MP-RAGE and reveal heterogeneity on the qMT maps, which is most pronounced in the unconstrained qMT maps.}
	\label{fig:InVivo_MS}
\end{figure*}

\begin{figure*}[htbp]
	\centering
	\ifOL
		\includegraphics[]{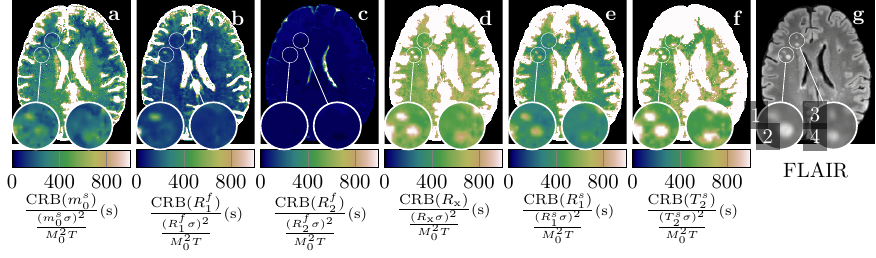}
	\else
		\input{Figures/In_Vivo_CRB.tex}
	\fi
	\caption{Cram\'er-Rao bound (CRB) of the in vivo parameter estimates shown in Fig. \ref{fig:InVivo_MS}a-f, normalized to resemble the inverse squared SNR of the parameter estimates (cf. Tab. \ref{tab:CRB}). We used the unconstrained MT model and the corresponding estimates for the CRB calculations. The FLAIR image is provided for anatomical reference.}
	\label{fig:InVivo_CRB}
\end{figure*}

\subsection{Precision of in vivo qMT estimates}
In order to gauge the precision of the qMT estimates, we calculated the CRB for each voxel in the transversal slice of an individual with MS (Fig.~\ref{fig:InVivo_CRB}). In WM, we find good agreement between the optimized CRB values (Tab.~\ref{tab:CRB}) and the CRB for the estimated parameters, confirming that the optimization for a single point in parameter space was adequate. The biggest deviations to Tab.~\ref{tab:CRB} can be found in $m_0^s$, which is slightly higher for the in vivo WM estimates, and in $R_1^f$, which is slightly lower for the in vivo WM estimates.

In the cortical GM, we find CRB values similar to the ones in WM for $m_0^s$, $R_1^f$, and $R_2^f$, and increased CRB values for $R_\text{x}$, $R_1^s$, and $T_2^s$, which is expected due to the reduced $m_0^s$. This effect is also in line with increased noise levels reported in Tab.~\ref{tab:ROImean} of the main paper and is even more pronounced in CSF, where $m_0^s$ is close to zero.
In the MS lesions, we see only slight increases in the CRB values $m_0^s$, $R_1^f$, and $R_2^f$, which gives confidence in the estimates of these parameters in lesions. The CRB values of $R_\text{x}$, $R_1^s$, and $T_2^s$, on the other hand, do increase MS lesions, in particular in lesions 1 and 2 (cf. Fig.~\ref{fig:InVivo_CRB}g).
This increase is likely related to the lesions' pronounced decrease in $m_0^s$.

\begin{figure}[htbp]
	\centering
	\ifOL
		\includegraphics[]{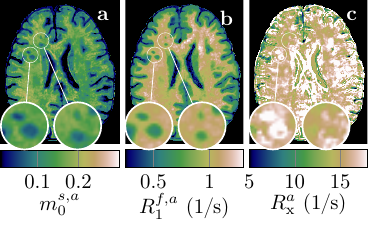}
	\else
		\begin{tikzpicture}[scale = 1, every text node part/.style={align=left}]
    \begin{scope}[spy using outlines={magnification=3, connect spies}]
        \begin{axis}[%
                width={2.0cm},
                height={2.0cm*1.214},
                axis on top,
                scale only axis,
                xmin=0,
                xmax=1,
                ymin=0,
                ymax=1,
                xtick = \empty,
                ytick = \empty,
                name=m0s,
                colormap name = gist_earth,
                colorbar horizontal,
                point meta min=0.01,
                point meta max=0.299,
                colorbar style={xlabel=$m_0^{s,a}$, height=0.3cm, yshift=0.2cm, xlabel style = {yshift = 0.15cm}, xticklabel style={/pgf/number format/fixed, /pgf/number format/precision=2}},
            ]
            \addplot graphics [xmin=0,xmax=1,ymin=0,ymax=1] {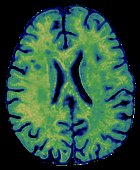};

            \coordinate (spypoint_s) at (axis cs: .25, .62);
            \coordinate (spyviewer_s) at (axis cs:0.2,0.125);
            \spy[circle, white,size=0.8cm] on (spypoint_s) in node at (spyviewer_s);

            \coordinate (spypoint_t) at (axis cs: .34, .72);
            \coordinate (spyviewer_t) at (axis cs:0.65,0.125);
            \spy[circle, white,size=0.8cm] on (spypoint_t) in node at (spyviewer_t);

            \node[text=white, anchor = north east] at (rel axis cs:  0.975,0.975) {\textbf{a}};
        \end{axis}
    \end{scope}

    \begin{scope}[spy using outlines={magnification=3, connect spies}]
        \begin{axis}[%
                width={2.0cm},
                height={2.0cm*1.214},
                axis on top,
                scale only axis,
                xmin=0,
                xmax=1,
                ymin=0,
                ymax=1,
                hide axis,
                name=R1f,
                at=(m0s.right of north east),
                anchor=north west,
                xshift = 0.1cm,
                colormap name = gist_earth,
                colorbar horizontal,
                point meta min=0.25,
                point meta max=1.3,
                colorbar style={xlabel=$R_1^{f,a}~(1/\text{s})$, height=0.3cm, yshift=0.2cm, xlabel style = {yshift = 0.15cm}, xticklabel style={/pgf/number format/fixed, /pgf/number format/precision=2}},
            ]
            \addplot graphics [xmin=0,xmax=1,ymin=0,ymax=1] {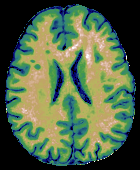};

            \coordinate (spypoint_s) at (axis cs: .25, .62);
            \coordinate (spyviewer_s) at (axis cs:0.2,0.125);
            \spy[circle, white,size=0.8cm] on (spypoint_s) in node at (spyviewer_s);

            \coordinate (spypoint_t) at (axis cs: .34, .72);
            \coordinate (spyviewer_t) at (axis cs:0.65,0.125);
            \spy[circle, white,size=0.8cm] on (spypoint_t) in node at (spyviewer_t);

            \node[text=white, anchor = north east] at (rel axis cs:  0.975,0.975) {\textbf{b}};
        \end{axis}
    \end{scope}

    \begin{scope}[spy using outlines={magnification=3, connect spies}]
        \begin{axis}[%
                width={2.0cm},
                height={2.0cm*1.214},
                axis on top,
                scale only axis,
                xmin=0,
                xmax=1,
                ymin=0,
                ymax=1,
                hide axis,
                name=R2f,
                at=(R1f.right of north east),
                anchor=north west,
                xshift = 0.1cm,
                align=center,
                colormap name = gist_earth,
                colorbar horizontal,
                point meta min=5,
                point meta max=18,
                colorbar style={xlabel=$R_\text{x}^a~(1/\text{s})$, height=0.3cm, yshift=0.2cm, xlabel style = {yshift = 0.15cm}, xticklabel style={/pgf/number format/fixed, /pgf/number format/precision=2}},
            ]
            \addplot graphics [xmin=0,xmax=1,ymin=0,ymax=1] {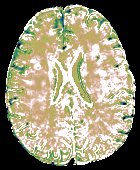};

            \coordinate (spypoint_s) at (axis cs: .25, .62);
            \coordinate (spyviewer_s) at (axis cs:0.2,0.125);
            \spy[circle, white,size=0.8cm] on (spypoint_s) in node at (spyviewer_s);

            \coordinate (spypoint_t) at (axis cs: .34, .72);
            \coordinate (spyviewer_t) at (axis cs:0.65,0.125);
            \spy[circle, white,size=0.8cm] on (spypoint_t) in node at (spyviewer_t);

            \node[text=white, anchor = north east] at (rel axis cs:  0.975,0.975) {\textbf{c}};
        \end{axis}
    \end{scope}
\end{tikzpicture}
	\fi
	\caption{Apparent quantitative MT maps when assuming $T_1^s = T_1^f$ in a participant with MS. The maps were calculated voxel-wise with Eqs.~\eqref{eq:R1a}--\eqref{eq:m0s_Taylor} and based on the maps depicted in Fig.~\ref{fig:InVivo_MS}. Note the different color scale in $R_1^{f,a}$ compared to Fig.~\ref{fig:InVivo_ctrl}.}
	\label{fig:InVivo_MS_R1a}
\end{figure}

\begin{figure}[htbp]
	\vspace{0.2cm}
	\centering
	\ifOL
		\includegraphics[]{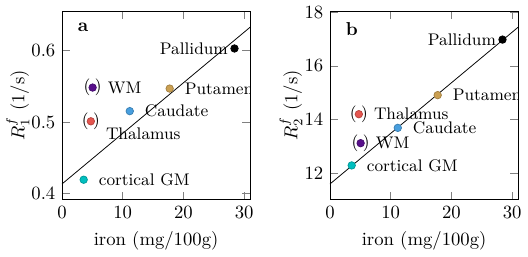}
	\else
		\begin{tikzpicture}[scale = 0.9]
    \begin{axis}[
            width=\columnwidth*0.4,
            height=\columnwidth*0.4,
            scale only axis,
            xmin = 0,
            xmax = 31,
            colormap={bw}{color(0cm)=(NYUpurple); color(1cm)=(turquois); color(2cm)=(TheLake); color(3cm)=(SpicyMustard); color(4cm)=(black); color(5cm)=(Pastrami)},
            xlabel={iron (mg/100g)},
            xticklabel style={/pgf/number format/fixed, /pgf/number format/precision=2},
            ylabel={$R_1^f$ (1/s)},
            ylabel style={yshift=-0.2cm},
            name=R1f,
        ]

        \addplot[only marks, scatter, scatter src=explicit]
        coordinates {
                (5.055e+00, 5.481e-01) [1]
                (3.583e+00, 4.194e-01) [2]
                (1.117e+01, 5.152e-01) [3]
                (1.774e+01, 5.467e-01) [4]
                (2.841e+01, 6.027e-01) [5]
                (4.760e+00, 5.010e-01) [6]
            };

        \node[anchor=west]       at (axis cs: 5.055e+00, 5.481e-01) {\small \, WM};
        \node[anchor=west]       at (axis cs: 3.583e+00, 4.194e-01) {\small \, cortical GM};
        \node[anchor=west]       at (axis cs: 1.117e+01, 5.152e-01) {\small \, Caudate};
        \node[anchor=west]       at (axis cs: 1.774e+01, 5.467e-01) {\small \, Putamen};
        \node[anchor=east]       at (axis cs: 2.841e+01, 6.027e-01) {\small Pallidum};
        \node[anchor=north west] at (axis cs: 4.760e+00, 5.010e-01) {\small \, Thalamus};

        \node at (axis cs: 5.055e+00, 5.481e-01)  {\small (\;)};
        \node at (axis cs: 4.760e+00, 5.010e-01)  {\small (\;)};

        \addplot[domain=0:31,samples=2, color=black] ({x}, {0.41336159192902133 + 0.007069779828355307 * x});

        \node[anchor=north west] at (rel axis cs: 0.05, .975)  {\textbf{a}};
    \end{axis}

    \begin{axis}[
        width=\columnwidth*0.4,
        height=\columnwidth*0.4,
        scale only axis,
        xmin = 0,
        xmax = 31,
        colormap={bw}{color(0cm)=(NYUpurple); color(1cm)=(turquois); color(2cm)=(TheLake); color(3cm)=(SpicyMustard); color(4cm)=(black); color(5cm)=(Pastrami)},
        xlabel={iron (mg/100g)},
        xticklabel style={/pgf/number format/fixed, /pgf/number format/precision=2},
        ylabel={$R_2^f$ (1/s)},
        ylabel style={yshift=-0.2cm},
        name=R2f,
        at=(R1f.east),
        anchor=west,
        xshift=1.5cm,
    ]

    \addplot[only marks, scatter, scatter src=explicit]
    coordinates {
            (5.055e+00, 1.312e+01) [1]
            (3.583e+00, 1.229e+01) [2]
            (1.117e+01, 1.369e+01) [3]
            (1.774e+01, 1.491e+01) [4]
            (2.841e+01, 1.698e+01) [5]
            (4.760e+00, 1.420e+01) [6]
        };

    \node[anchor=west] at (axis cs: 5.055e+00, 1.312e+01) {\small \, WM};
    \node[anchor=west] at (axis cs: 3.583e+00, 1.229e+01) {\small \, cortical GM};
    \node[anchor=west] at (axis cs: 1.117e+01, 1.369e+01) {\small \, Caudate};
    \node[anchor=west] at (axis cs: 1.774e+01, 1.491e+01) {\small \, Putamen};
    \node[anchor=east] at (axis cs: 2.841e+01, 1.698e+01) {\small Pallidum};
    \node[anchor=west] at (axis cs: 4.760e+00, 1.420e+01) {\small \, Thalamus};

    \node at (axis cs: 5.055e+00, 1.312e+01)  {\small (\;)};
    \node at (axis cs: 4.760e+00, 1.420e+01)  {\small (\;)};

    \addplot[domain=0:31,samples=2, color=black] ({x}, {11.596269888495543 + 0.18867537062956058 * x});

    \node[anchor=north west] at (rel axis cs: 0.05, .975)  {\textbf{b}};
\end{axis}
\end{tikzpicture}
	\fi
	\vspace{-0.5cm}
	\caption{Relaxation rates, measured with our qMT approach, as a function of iron concentrations, taken from the literature \citep{Hallgren.1958}. \textbf{a:} Except for WM and the thalamus, which were excluded from the fit, the longitudinal relaxation rate $R_1^f = 1/T_1^f$ follows in good approximation a linear function ($R^2 = 0.94$).
		\textbf{b:} The transversal relaxation rate of the free pool $R_2^f = 1/T_2^f$ is even better described by a linear model ($R^2 = 0.9998$).
	}
	\label{fig:Iron}
\end{figure}

\begin{figure}[htbp]
	\centering
	\ifOL
		\includegraphics[]{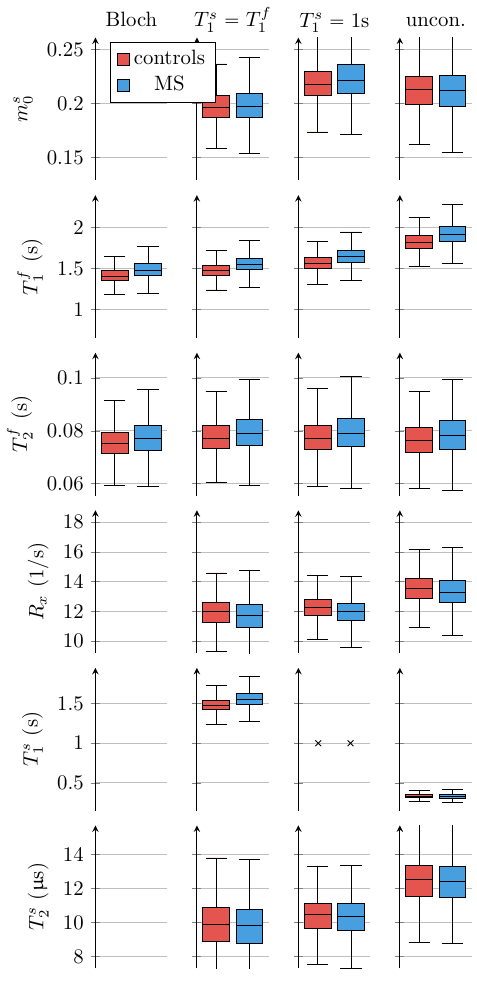}
	\else
		\input{Figures/ROI_cf_Models.tex}
	\fi
	\caption{Comparison of the parameter estimates between a Bloch model, two traditional MT models that assume $T_1^s = T_1^f$ and $T_1^s = 1$s, respectively, and the proposed unconstrained MT model. This analysis pools all normal-appearing white matter voxels of 4 healthy participants and 4 individuals with MS, respectively. Note that the $T_1^s = T_1^f$ column depicts the same longitudinal relaxation time estimates twice to illustrate the differences to the unconstrained MT model.
	}
	\label{fig:ROI_cf_Models}
\end{figure}

\begin{figure}[htbp]
	\centering
	\ifOL
		\includegraphics[]{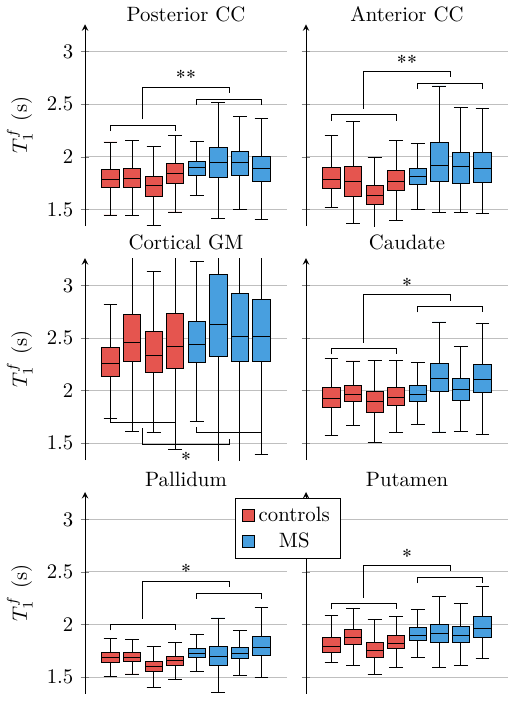}
	\else
		\input{Figures/ROI_R1f.tex}
	\fi
	\caption{ROI analysis of the unconstrained model's $T_1^f$. For all six WM and GM ROIs, we found statistically significant differences between participants with MS and controls. The markers $*$ and $**$ indicate statistically significant differences at the $p < 0.05$ and $p < 0.01$ levels. }
	\label{fig:ROI_R1f}
\end{figure}

\bibliographystyle{elsarticle-harv}
\bibliography{library}%

\makeatletter \@input{MT_T1_Paper.aux.tex} \makeatother